\documentclass{aa} 
\usepackage{natbib}
\usepackage{graphicx}
\usepackage{ulem}
\usepackage[varg]{txfonts}
\usepackage{natbib}
\usepackage[colorlinks=true,citecolor=blue,linkcolor=blue]{hyperref}

\usepackage[dvipsnames]{xcolor}
\usepackage{soul}

\begin{document}

\title{Dimming GRS~1915+105 observed with NICER and Insight--HXMT}
\titlerunning{Dimming GRS~1915+105 observed with NICER and Insight--HXMT}

\author{
M.~Zhou\inst{1}\and
V.~Grinberg\inst{2}\and
A.~Santangelo\inst{1}\and
C.~Bambi\inst{3,4}\and
Q.~Bu\inst{5}\and
C.~M.~Diez\inst{6}\and
L.~Kong\inst{1}\and
J.~F.~Steiner\inst{7}\and
Y.~Tuo\inst{1}
}

\institute{Institut f\"ur Astronomie und Astrophysik, Universität T\"ubingen, Sand 1, 72076 T\"ubingen, Germany \\
  \texttt{menglei.zhou@astro.uni-tuebingen.de}
\and European Space Agency (ESA), European Space Research and Technology Centre (ESTEC), Keplerlaan 1, 2201 AZ Noordwijk, The Netherlands 
\and Center for Astronomy and Astrophysics, Center for Field Theory and Particle Physics, and Department of Physics, Fudan University, Shanghai 200438, China
\and School of Natural Sciences and Humanities, New Uzbekistan University, Tashkent 100007, Uzbekistan
\and Institute of Astrophysics, Central China Normal University, Wuhan 430079, China
\and European Space Astronomy Centre (ESAC), Camino Bajo del Castillo s/n, Villanueva de la Cañada, 28692 Madrid, Spain
\and Harvard-Smithsonian Center for Astrophysics, 60 Garden Street, Cambridge, MA 02138, USA
}

\date{ -- / --}

\abstract{The black hole X-ray binary GRS~1915+105 was bright for 26 years since its discovery and is well known for its disk instabilities, quasi-periodic oscillations, and disk wind signatures. We report a long-term spectral-timing tracing of this source from mid-2017 until the onset of the so-called obscured state based on the complete data from the Neutron Star Interior Composition Explorer (NICER) and the Insight–-Hard X-ray Modulation Telescope (HXMT), whose hard coverage decisively informs the modeling at lower energies. In the soft state predating 2018, we observed highly ionized winds. However, in the hard state shortly before transitioning into the obscured state on May 14, 2019 (MJD\,58617), the winds exhibited a discernible reduction in ionization degree ($\log \xi$), which decreased from above 4 to approximately $3$. Our analysis involves the measurement of the frequencies of the quasi-periodic oscillations and the estimation of the properties of the ionized winds and the intensities of different spectral components through spectroscopy during the decay phase. We studied the origin of these infrequently observed warm outflows in the hard state. The launching radius of the winds in the hard decay phase is similar to that in the soft state, which indicates that the launching mechanism of these winds likely is the same in both states. The presence of the ionized winds is preferentially dependent on the periphery of the accretion disk, but it is not directly related to the corona activities in the center of the binary system. 
}


    \maketitle

\section{Introduction}\label{Sect:intro}
Black hole binaries (BHBs) consist of a black hole and a donor star companion~\citep[for a review on BH binaries, see e.g.][and the references therein]{Remillard_2006}. The black hole (BH) accretes matter from its companion star. The accretion process leads to the release of intense radiation and outflows that can be observed across the electromagnetic spectrum. GRS~1915+105 is a particularly interesting BHB. First discovered by the Gamma-Ray Astronomical Telescope / Wide Angle Telescope for Cosmic Hard X-rays (GRANAT/WATCH) all-sky monitor in 1992~\citep{Castro-Tirado_1992, Castro-Tirado_1994}, GRS~1915+105 has been well studied since. The captivating phenomena of this source, such as the exhibition of apparently superluminal radio jets~\citep{Mirabel_1994}, rarely seen high-frequency quasi-periodic oscillation (HFQPO) features in the timing domain~\citep{Morgan_1997}, the highly ionized winds generated from the accretion disk~\citep{Lee_2002, Ueda_2009}, and the characteristic patterns of its light curves that can be classified into at least 14 different classes~\citep{Belloni_2000, Klein-wolt_2002, Hannikainen_2005, Zoghbi_2016, Athulya_2022, Shi_2023}, all make GRS~1915+105 valuable and important for our understanding of astrophysical processes such as accretion and the outflow formation, including the jet ejection and disk winds. 

Unlike other transient X-ray binaries that brighten and fade during weeks or months~\citep[e.g.,][]{Remillard_1999, Homan_2001, Williams_2020}, GRS~1915+105 was consistently bright for 26 years since its discovery and was thought to be a quasi-persistent source. However, its luminosity unexpectedly underwent an exponential decay in early 2018~\citep{Negoro_2018_Atel11828}. During this flux decline, the source also showed prominent and narrow low-frequency quasi-periodic oscillation (LFQPO) features~\citep{Koljonen_2021}. It was then thought that GRS~1915+105 had nearly completed its 26-year-long outburst and entered its quiescent state. In April 2019 ($\sim$\,$\text{MJD}\, 58600$), the source became dimmer during a pre-flare dip~\citep{Homan_2019_Atel12742}. A few days later, short flares lasting for hours were observed in the radio and X-ray band~\citep{Motta_2019_Atel12773, Motta_2021, Koljonen_2019_Atel12839}, but they had an even lower average X-ray flux. The LFQPO features within 1--10\,Hz were either halted or became undetectable synchronously. However, during a 60-day rebrightening phase in mid-2021, strong QPO features with an unprecedentedly low frequency ($f_\text{QPO} \simeq 0.2$\,Hz) were discovered~\citep{Kong_2024}. Intriguingly, despite the rapid luminosity decay measured in X-rays, the radio observations suggest that GRS~1915+105 has remained active in the accretion process after mid-2019. 

Generally, the physics of the state transition of BHBs is not fully understood. Empirically, a typical outburst observed in a BHB can form a \texttt{q}-track in its hardness-intensity diagram (HID)~\citep{Homan_2001, Fender_2006}. During the hard state in which the (cutoff) power-law emission dominates in X-rays, jet emission is observed in radio and infrared bands, and LFQPOs in the timing domain often emerge. In contrast, in the soft state when thermal radiation becomes primary, the jet typically becomes either weak and unstable or undetectable, with a power spectrum dominated by flicker noise. The central frequencies of the LFQPOs are found to have a negative correlation with the spectral hardness up to at least 90\,keV~\citep[e.g.,][]{Zdziarski_2004, Soleri_2008, Zhou_2022}. As an ingredient involved in the accretion process, ionized disk winds are thought to play an important role during the state transition and are preferentially discovered in the soft state, where the jet emission is usually quenched~\citep{Neilsen_2009, Ponti_2012}. The state-dependent anticorrelation between the disk winds and the jets indicates that the disk winds may work as an alternative mode to eject matters. Previous studies have shown that GRS~1915+105 displays prominent blueshifted ionization absorption features, for instance, the Fe\,{\sc xxv} He$\alpha$ line and the Fe\,{\sc xxvi} Ly$\alpha$ line, which are located at 6.7\,keV and 7\,keV, respectively~\citep{Ueda_2009, Miller_2016, Neilsen_2018}. 

It is still unclear which mechanism drives the ionized absorption outflow. It can be generated by radiation pressure if the luminosity of the source approaches the Eddington limit~\citep{Proga_2002, Higginbottom_2015}, by thermal pressure when the gas of the disk atmosphere is irradiated by the energetic photons from the inner part of the accretion disk~\citep{Begelman_1983, Woods_1996, Done_2018}, or by magnetic pressure given the strong magnetic field close to the BH~\citep{Fukumura_2017}. Additionally, the partial concealment of cold dense absorption detected in some BHBs may also arise from an internal obscured clump that failed to escape from the strong gravitational field~\citep{Miller_2020}. The presence of the obscured clump, which exhibits fast absorption variabilities that can be seen in BHBs~\citep[see e.g.,][]{Motta_2017_Oct, Motta_2017_Jun, Koljonen_2020, Balakrishnan_2021} and active galactic nuclei (AGNs)~\citep[e.g.,][]{Matt_2003}, provides significant diagnostic information for understanding how accretion behaves within the full range of BH masses from a few $M_\odot$ to $10^9\, M_\odot$. 

We focus on the period just before GRS~1915+105 entered the obscured state and use all the public data that were observed between MJD\,57932 (June 8, 2017) and MJD\,58608 (May 5, 2019). The data were obtained from the archive of X-ray Timing Instrument (XTI) of the Neutron Star Interior Composition Explorer (NICER; \citealt{Gendreau_2016}) and Insight–-Hard X-ray Modulation Telescope (Insight--HXMT or HXMT; \citealt{Zhang_2014, Zhang_2020}). We perform a spectral-timing analysis based on these data by tracing the evolution of key parameters and probing their correlations using the simultaneous broadband coverage by both instruments, in particular, the high-energy coverage afforded by HXMT where available, to also inform the single-instrument fits. The results contribute to our understanding of the physics of the accretion process and the mechanisms behind the state transitions of transient BHBs. The remainder of this paper is structured as follows: In Sect.~\ref{Sect:obs} we present the long-term behaviors of the source and an overview of the data we analyze in this paper. The spectral-timing analysis is detailed in Sect.~\ref{Sect:spec-timing} and the results are presented in Sect.~\ref{Sect:results}. We discuss the results of our analysis in Sect.~\ref{Sect:discussion}. A summary and outlook are presented in Sect.~\ref{Sect:conclusion}. 

\section{Observations and data reduction}\label{Sect:obs}
\begin{figure*}
    \centering
    \includegraphics[width = 0.99\textwidth]{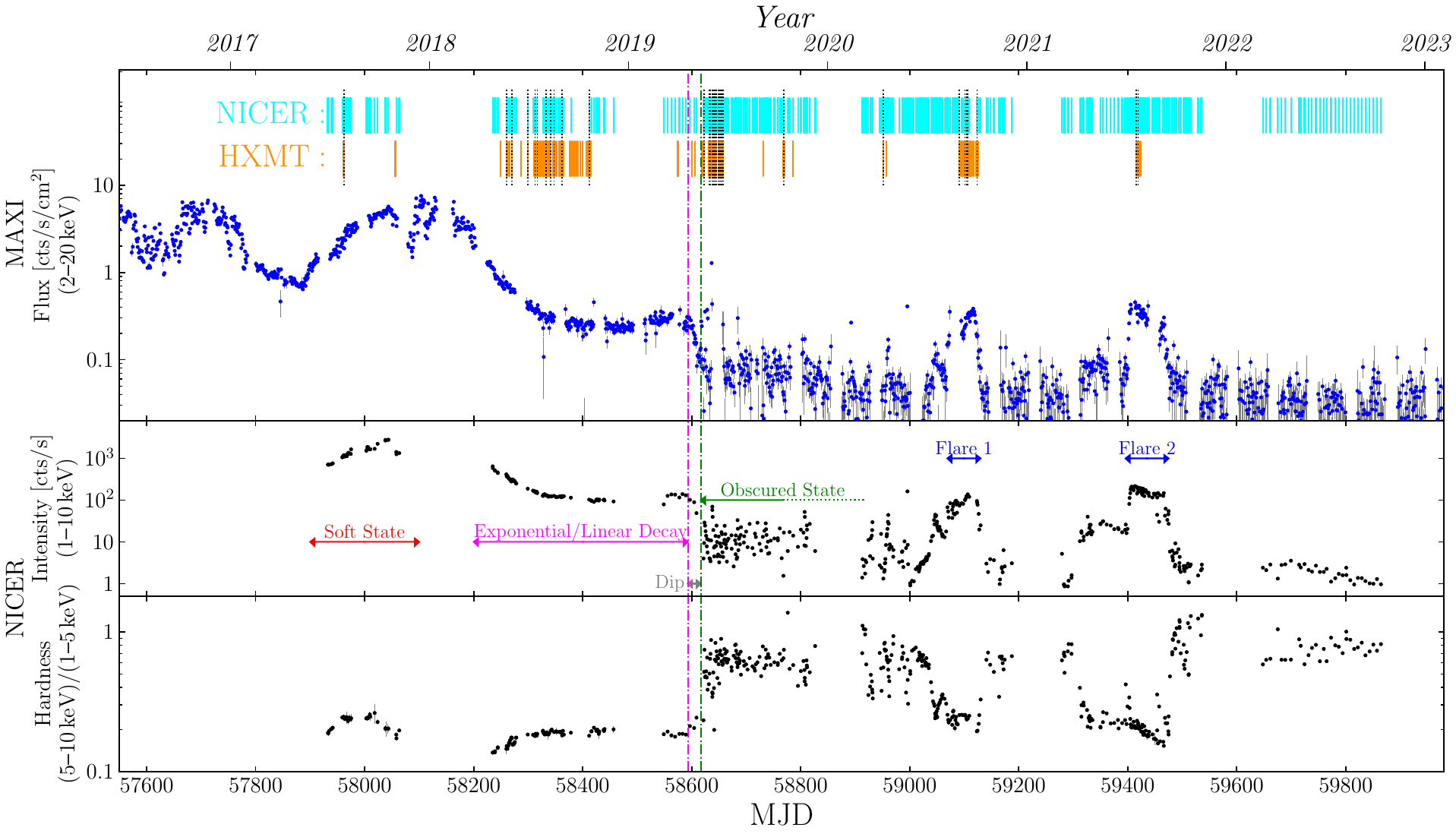}
    \caption{Long-term behavior of GRS~1915+105 monitored by MAXI and NICER. Top panel: Daily flux measured by MAXI in the 2--20\,keV band. The duration of NICER and Insight--HXMT observations is shown at the top with cyan and dark orange stripes. The dotted black lines denote the available quasi-simultaneous observations by two instruments. Middle panel: Intensity of GRS~1915+105, defined by the raw photon count rate registered in the 1--10\,keV band of NICER. The spectral states are classified with spectral-timing features and denoted with arrows. More details can be found in the literature. Bottom panel: Hardness of the spectra, defined by the raw count rate between the 5--10\,keV band and the 1--5\,keV band of NICER. The vertical dash-dotted magenta line ($\sim$ MJD\,58593) represents the start point of the pre-flare dip, and the dash-dotted green line ($\sim$ MJD\,58617) marks the passing of this source into the obscured state. }
    \label{f-longterm_lc}
\end{figure*}

The long-term behavior of GRS~1915+105 since 2017 is shown in Fig.~\ref{f-longterm_lc}. The light curve monitored by MAXI~\citep{Matsuoka_2009} indicates that this source was consistently active before 2018, but then underwent a transition from an exponential decay to a linear decay until the beginning of May 2019 (we followed the state definitions of \citealt{Koljonen_2021}). After this, the source became extraordinarily faint, and its flux was reduced by a factor of 10. It exhibited three month-long flares and several short flares that only lasted a few hours in the X-ray band~\citep[see e.g.,][]{Neilsen_2020, Kong_2021}, but it remained particularly active in the radio band~\citep{Motta_2021}. The abundant observations focusing on GRS~1915+105 by NICER and HXMT allowed us to closely trace the evolution of this source after 2017. We are particularly interested in the observations before the source entered the purely obscured state, and we used all the available NICER and HXMT data before MJD\,58617. 

The NICER and the HXMT both produce essential spectral-timing data. The XTI of NICER consists of an aligned collection of 56 X-ray concentrator optics and silicon drift-detector pairs that extend to a softer X-ray band (0.25--10\,keV) with a time-tagging resolution of $\sim\,$100 nanoseconds. HXMT instead has a broader energy band that nominally ranges from 2 to around 250\,keV, which is realized by three collimated telescopes: the Low-Energy (LE) telescope, whose energy range covers the 1--15\,keV band~\citep{Chen_2020_le}; the Medium Energy (ME) telescope, covering 5--30\,keV~\citep{Cao_2020_me}; and the High Energy (HE) telescope, covering the 25--250\,keV band~\citep{Liu_2020_he}. In particular, the fast temporal response and large effective area of HXMT at higher energies allowed us to conduct spectral-timing studies above 15\,keV, and in particular, to better constrain the continuum spectral models by providing high-energy leverage on the spectral shape. We present the details of the data processing in the following subsections. 

\subsection{NICER data}
We processed all NICER data using the tools provided in \texttt{HEASOFT} v6.32.1 and the calibration database \texttt{xti20221001}. The cleaned event lists are produced by the standard \texttt{nicerl2} task, including calibration and filtering steps. To minimize the background effects, we applied a custom set of selection criteria to identify good time intervals (GTIs) based on standard NICER screening criteria: an overshoot rate lower than 10\,cts per FPM (\texttt{FPM\_OVERONLY\_COUNT<10}), an undershoot rate lower than 200\,cts per FPM (\texttt{FPM\_UNDERONLY\_COUNT<200}), and a geomagnetic cutoff-rigidity greater than 2\,GeV (\texttt{COR\_SAX>2}). The photons registered by the noisy detectors (\texttt{DET\_ID=14,34,43,54}) were also omitted~\citep[e.g.,][]{Bogdanov_2019}. Finally, we extracted the spectra and light curves within our selected GTIs with \texttt{nicerl3-spect} and \texttt{nicerl3-lc}. The background spectrum was estimated with model \texttt{3C50}~\citep{Remillard_2022}. The energy band for the spectral analyses is 0.4--10\,keV. 

\subsection{HXMT data}
For the HXMT data, we used the pipeline based on the Insight--HXMT Data Analysis Software (\texttt{HXMTDAS}) v2.05 and the calibration database v2.06. The GTIs were selected according to the following criteria: an elevation angle larger than $10^\circ$ (\texttt{ELV>10}), a geomagnetic cut-off rigidity higher than 8\,GeV (\texttt{COR>8}), an offset angle from the pointing direction smaller than $0.04^\circ$ (\texttt{ANG\_DIST<0.04}), and at least 300 seconds before and after the South Atlantic Anomaly passage (\texttt{T\_SAA>=300 \&\& TN\_SAA>=300}). We adopted an additional criterion for the LE telescope: the elevation angle for the bright Earth must be larger than $30^\circ$. Since the individual detectors of the HXMT instruments have different fields of view (FoVs), we selected photon events that were registered by detectors with a small FoV for the following analysis. 

The energy bands for the spectral analyses were 2--10\,keV for LE, 10--30\,keV for ME, and 28--90\,keV for HE. We additionally ignored photons within the 21--24\,keV band of the ME telescope to avoid detecting the silver K-shell fluorescent lines produced by the material of the instrument~\citep{Li_2020_flight}. We used the tools \texttt{LEBKGMAP}~\citep{Liao_2020_lebackground}, \texttt{MEBKGMAP}~\citep{Guo_2020_mebackground}, and \texttt{HEBKGMAP}~\citep{Liao_2020_hebackground} that are provided by the HXMT team to estimate the instrumental background. 

\subsection{Joint fitting with NICER and HXMT data}\label{Sect:obs:joint}

NICER has a large effective area in the soft X-ray band ($\sim$\,$1900\, \mathrm{cm^{2}}$, peaking at 1.5\,keV), which makes it well suited for probing absorption characteristics, while HXMT covers a wider energy range up to roughly 250\,keV, which enables a good constraint on the slope of the continuum. 
We exploited the combination of the two X-ray missions where simultaneous data are available and used the results to inform models where the coverage exist only below 10\,keV. This allowed an improved modeling and better model constraints.
%To accomplish this, we choose observations from NICER and HXMT that possess time overlap with each other. 
A total of 20 joint spectra are available before the obscured state of GRS~1915+105 (see the top panel in Fig.~\ref{f-longterm_lc}). Seven of these occurred before 2018 (in the soft state) and 13 during the exponential and linear decay, during which the spectra are relatively hard. We optimized our modeling by conducting a joint fitting, which helped us to rule out unsound parameters and obtain better constraints on various theories. We used NICER/XTI and HXMT/ME together to extend our spectra to higher energies for observations prior to 2018, as HXMT/HE is predominantly affected by background. During the decay phase, we combined the spectra from NICER/XTI, HXMT/ME, and HXMT/HE. 

\section{Spectral-timing analysis}\label{Sect:spec-timing}
We used \texttt{ISIS} v1.6.2~\citep{Houck_2000}, which allowed us access to the defined models in \texttt{Xspec}~\citep{Arnaud_1996}, to perform the spectral-timing fitting. 
\subsection{Timing analysis}\label{Sect:spec-timing:timing}
All the light curves obtained from both NICER and HXMT were directly generated from the corresponding screened event lists and were processed with \texttt{Stingray}~\citep{Bachetti_2021, Huppenkothen_2019_stingrayb, Huppenkothen_2019_stingraya}. The selected energy range for NICER/XTI is 1--10\,keV. The time resolution ($\Delta t$) of these light curves was set to $2^{-9}\,\text{s} \approx 2\,\text{ms}$. The power spectrum densities (PSDs) were computed using segments with a length of $2^{14} \times \Delta t = 32\,\text{s}$. Thus, the Nyquist frequency for these PSDs is $f_{\text{max}} = 1/(2 \Delta t) = 256 \, \text{Hz}$, and the lowest frequency we were able to access is $f_{\text{min}} = 1/(n_{\text{bins}} \Delta t) = 0.03125 \, \text{Hz}$. The frequencies $f$ were rebinned logarithmically between $f_{\text{min}}$ and $f_{\text{max}}$, with an equally spaced grid in logarithmic scale $\mathrm{d}f/f = 0.03$. The expectations of the white noise were reduced according to \citet{Zhang_1995}. We omitted the data above 30\,Hz since they are dominated by Poisson noise. 

\begin{figure}
    \centering
    \includegraphics[width = 0.49\textwidth]{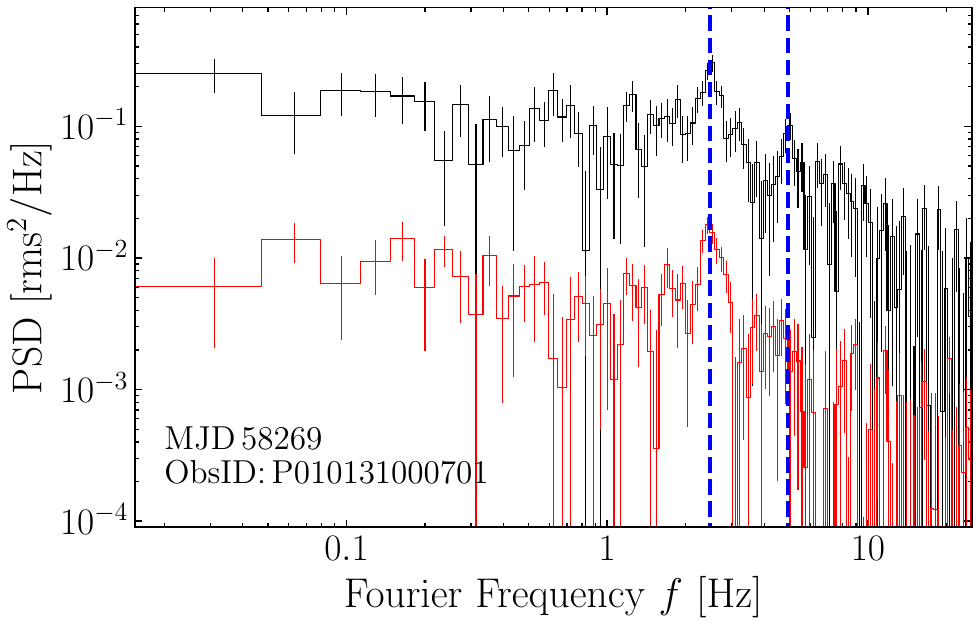}
    \caption{Typical PSDs simultaneously observed with HXMT/LE (in black) and HXMT/ME (in red). The PSDs of LE are multiplied by a constant of 20 to avoid overlapping in the figure. The central frequencies of the main QPO and its first harmonic are 2.47\,Hz and 4.94\,Hz, respectively, as indicated by two dashed blue lines. }
    \label{f-psdhxmt}
\end{figure}

The spectral states of GRS~1915+105 can clearly be classified by the timing properties. Before 2018, GRS~1915+105 was in a spectrally soft state, and its light curve is predominantly categorized as class $\lambda$~\citep{Belloni_2000, Neilsen_2018}. The flicker noise dominated the power spectrum, without unambiguous LFQPO features in the 1--10\,Hz band. From the end of April 2018 to the end of April 2019, the power spectra exhibited at least two prominent LFQPO features whose central frequencies were in the 1--10\,Hz range. At the end of the decay phase, the power spectra were dominated by Poisson noise since MJD\,58621 (May 18, 2019), marking the onset of the obscured state. During the year before the onset of the obscured state, we additionally examined the data from HXMT/LE and HXMT/ME. We found that the central frequencies of the LFQPO components are consistent in these simultaneous observations (see, e.g., Fig.~\ref{f-psdhxmt}), even though the PSDs produced by HXMT are noisier than those of NICER because HXMT has stronger background effects. The QPOs thus clearly extend above 10\,keV, that is, in a range that is dominated by the continuum with no disk contribution.  

\begin{figure}
    \centering
    \includegraphics[width = 0.49\textwidth]{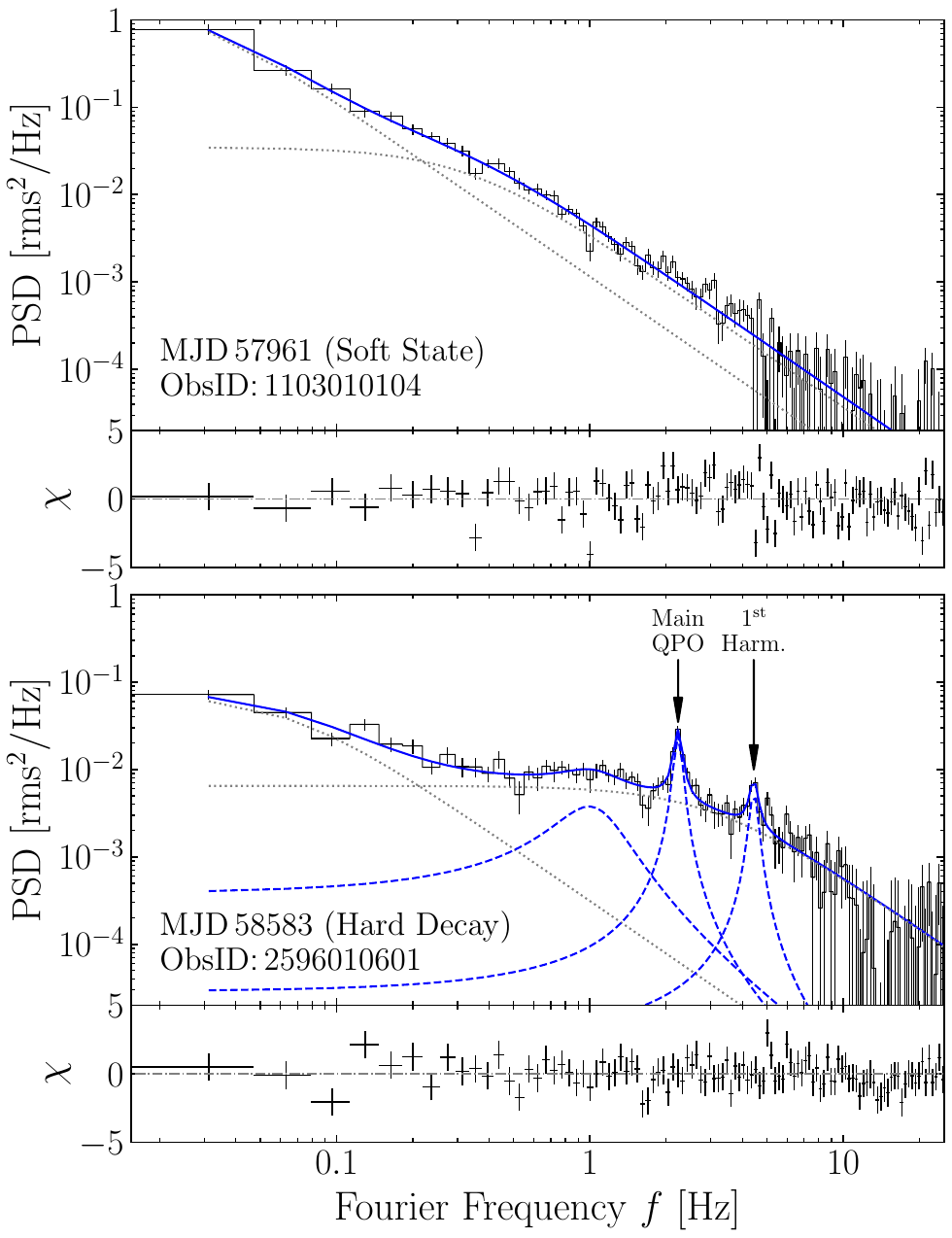}
    \caption{Two typical power spectra produced by NICER data in the soft (upper panel) and hard state during the decay phase (lower panel) with their best-fit models. The Lorentzians with a central frequency at zero are indicated with dotted gray lines, and the Lorentzians with a nonvanishing central frequency are indicated with dashed blue lines. As the best-fit model, the sums of all the Lorentzians are denoted with solid blue lines. The measured central frequencies of the main QPO and the first harmonic in the hard state are $2.23^{+0.18}_{-0.19}\,\mathrm{Hz}$ and $4.44^{+0.05}_{-0.06}\,\mathrm{Hz}$, respectively. }
    \label{f-psdfit}
\end{figure}

\begin{figure}
    \centering
    \includegraphics[width = 0.49\textwidth]{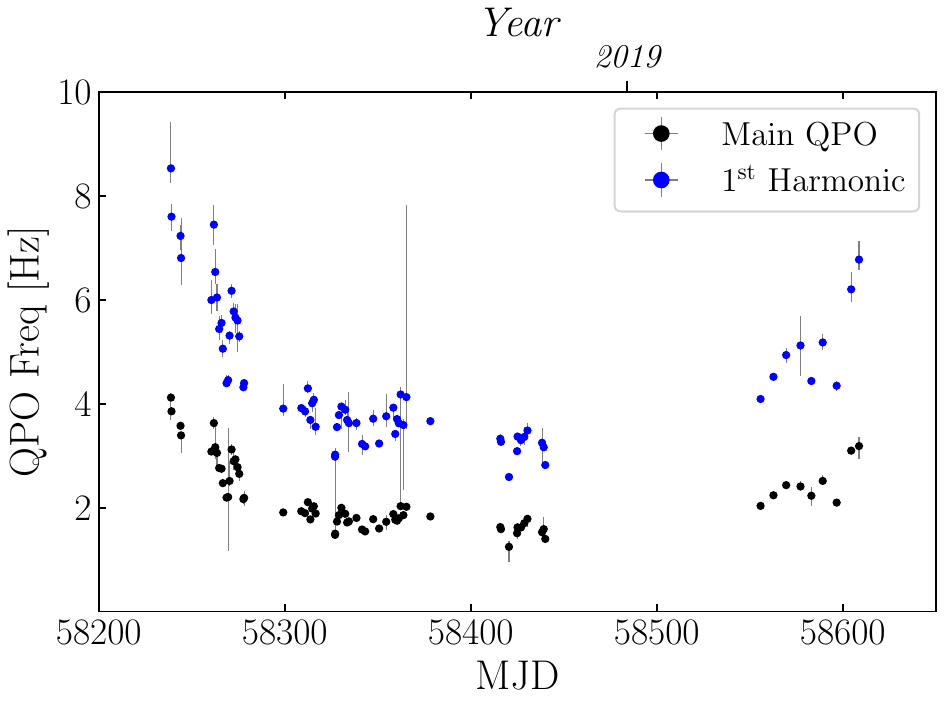}
    \caption{Evolution of the QPO frequencies vs. time during the exponential and linear decay. }
    \label{f-qpo-mjd}
\end{figure}

We modeled the NICER spectra because they are less affected by noise. To do this, we chose an empirical model consisting of two Lorentzian functions with central frequencies fixed at zero to characterize the noise continuum ~\citep{Nowak_2000}. The combination of these two Lorentzians is appropriate for flicker and flat-top noise with a high-frequency cutoff. The LFQPO signals were captured by three additional Lorentzian functions with free central frequencies. Two typical fits with this model are shown in Fig.~\ref{f-psdfit}. We obtained a relatively good fit for most of our power spectra. The most significant Lorentzian signal with a lower frequency is identified as the main QPO component, and the second with a higher frequency is recognized as the first harmonic. We show their central frequencies along with their posterior uncertainties versus time in Fig.~\ref{f-qpo-mjd}, and the correlation between these two components is plotted in the upper panel of Fig.~\ref{f-qpo-freq}. Our findings suggest that there is always a 2:1 relation between the central frequencies of two QPO components, which measures LFQPO frequencies more soundly. While we occasionally observe a third QPO component, there are no clear patterns in its appearance concerning the former two prominent QPO features. 

\subsection{Spectral analysis}\label{Sect:spec-timing:spectral}
We adopted a systematic error of 0.5\% and 1.0\% for HXMT/ME and HXMT/HE, respectively~\citep{Li_2020_flight}, and of 1.5\% for NICER/XTI\footnote{\url{https://heasarc.gsfc.nasa.gov/docs/nicer/analysis_threads/cal-recommend/}}. All the spectra were grouped using the optimal binning method detailed in \citet{Kaastra_2016}, and we ensured that there were 30 photon counts per bin at least. 

\subsubsection{Preliminary modeling for {NICER} only}\label{Sect:spec-timing:spectral:pre}
\begin{figure}
    \centering
    \includegraphics[width = 0.49\textwidth]{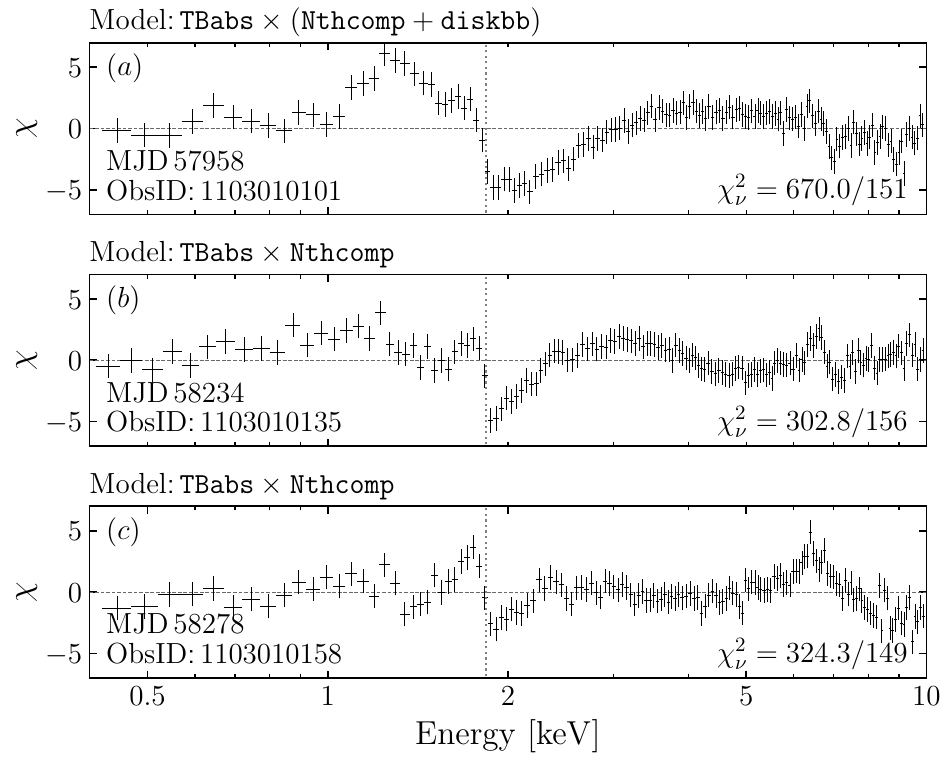}
    \includegraphics[width = 0.49\textwidth]{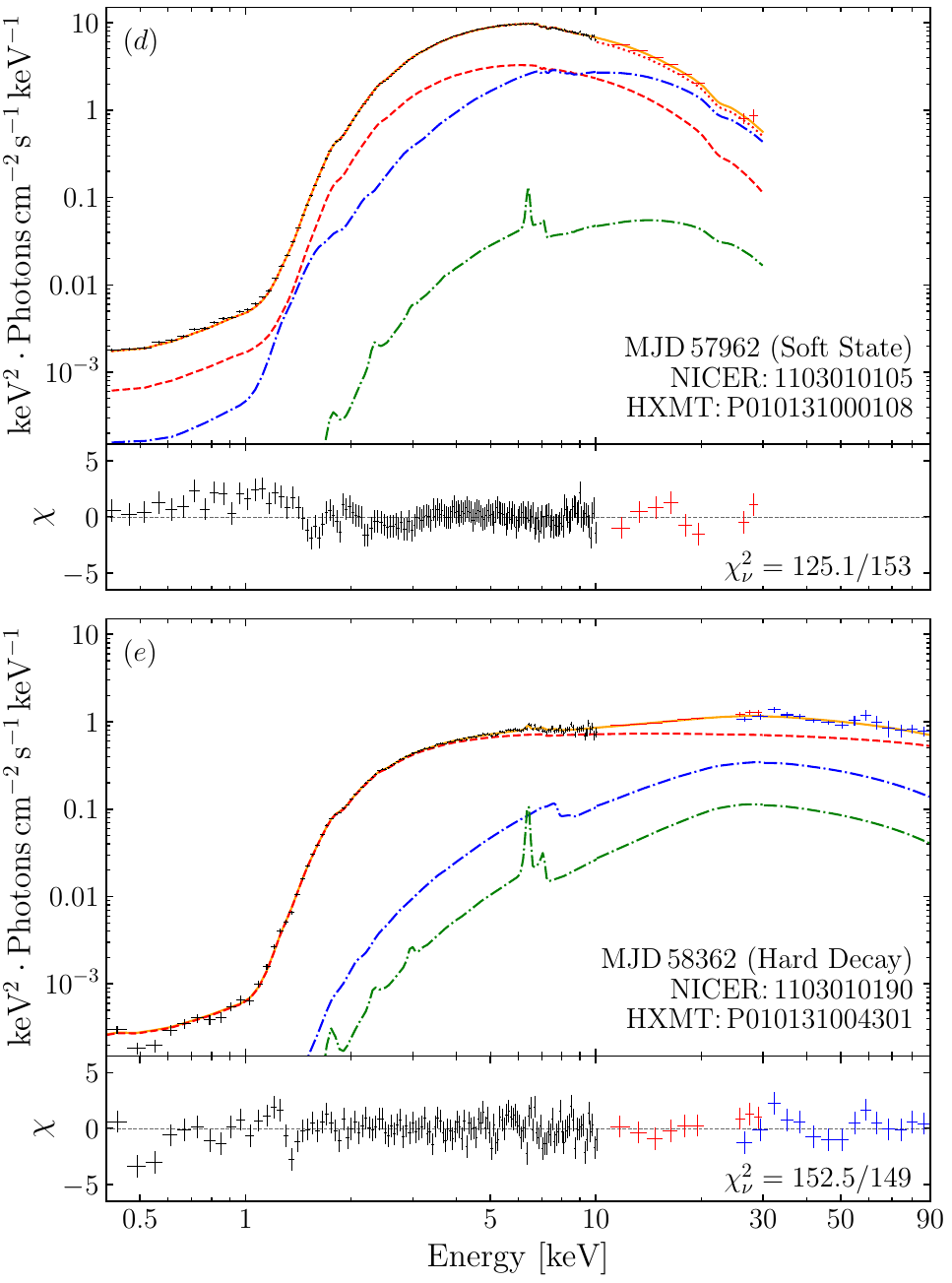}
    \caption{The residuals and the spectral composition diagrams for different states. From (a) to (c): Residuals for the preliminary modeling to fit the NICER observation during the soft state (MJD\,57958) at the beginning and the middle of the decay phase (MJD\,58234 \& 58278). An additional silicon absorption edge (1.84\,keV, denoted by the dotted line) and a narrow iron K$\alpha$ line with a cutoff energy lower than 7\,keV are plotted as well. From (d) to (e): Two typical merged unfolded spectra of GRS~1915+105 in different states, fit by the physical model illustrated in Eq.~\ref{eq-model}. Different spectral components, i.e., the power-law continuum, the relativistic reflection component from the accretion disk, and the nonrelativistic reflection component from a distant reflector, are shown with the dashed red line, the dash-dotted blue line, and the dash-dotted line line, respectively. See the text for detailed information. }
    \label{f-res}
\end{figure}

\begin{figure}
    \centering
    \includegraphics[width = 0.49\textwidth]{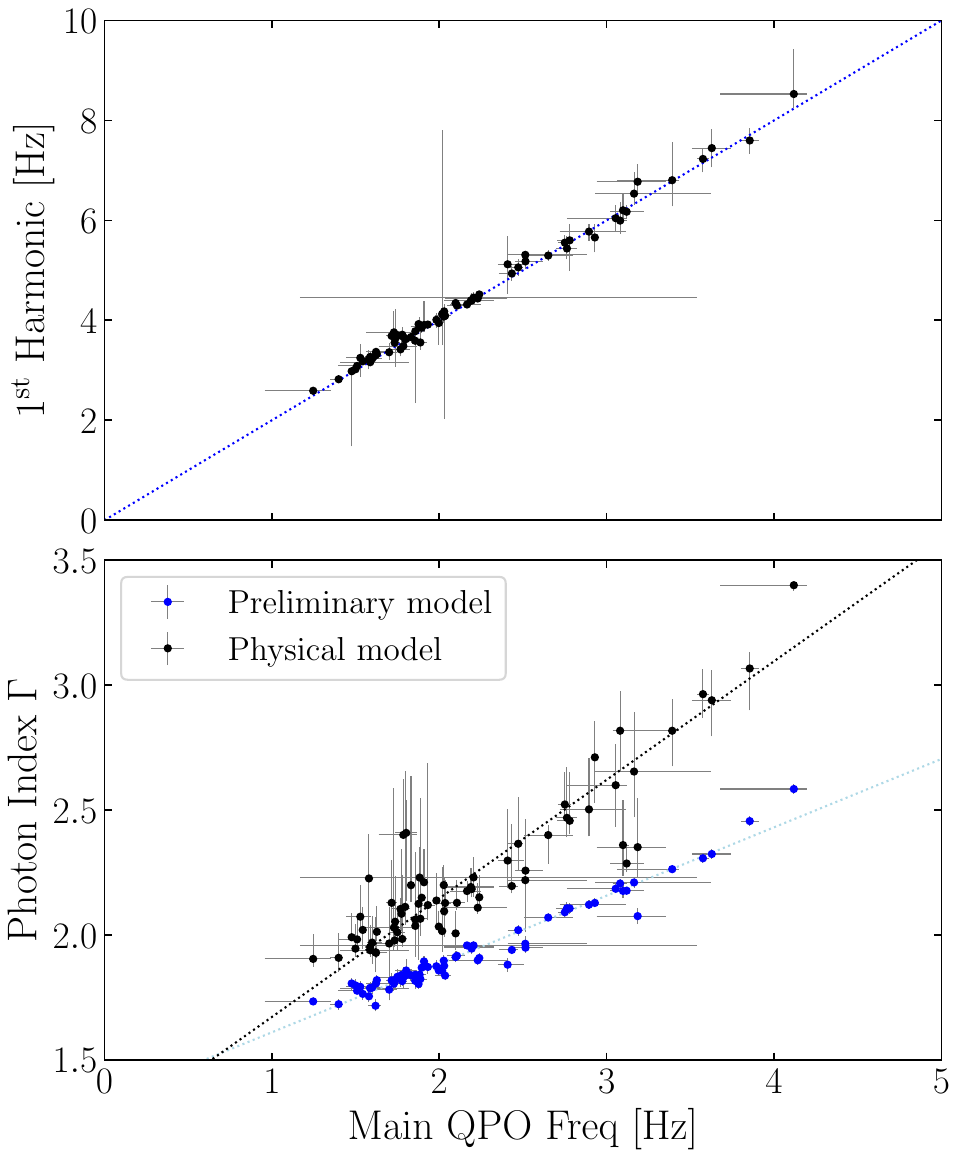}
    \caption{Upper panel: Correlation between the main QPO frequency and the first harmonic. The dotted blue line indicates the double relation. Lower panel: Correlation between the main QPO frequency and the measured photon index $\Gamma$ with the preliminary model and an improved physical model described in Sects.~\ref{Sect:spec-timing:spectral:pre} and \ref{Sect:spec-timing:spectral:physical}, respectively. The dotted lines show the best fits assuming a linear model. }
    \label{f-qpo-freq}
\end{figure}

We first tried to fit the NICER spectra with a simple continuum of the Comptonization emission produced by the corona (\texttt{Nthcomp}; \citealt{Zdziarski_1996, Zycki_1999}), multiplied by the Galactic absorption model \texttt{TBabs} to characterize the photoelectric absorption by the interstellar medium (ISM). The cross sections were provided by \citet{Verner_1996} and the element abundances were given by \citet{Wilms_2000}. The multitemperature thermal radiation generated from the accretion disk (\texttt{diskbb}; \citealt{Mitsuda_1984, Makishima_1986}) was further included in our model in the soft state. Typical residuals are presented in the upper panel of Fig.~\ref{f-res}. These exploratory fits are usually poor, with high reduced $\chi^2$s, but we obtained a well-constrained estimate of the photon index of the continuum. The correlation of the main QPO frequencies and the measured photon index $\Gamma$ is shown in blue in the lower panel of Fig.~\ref{f-qpo-freq}. A quasi-linear trend between these two parameters is observed even with this preliminary model. 

In the soft state, we detected absorption at $\sim$ 7\,keV, which we identified with Fe {\sc xxvi} Ly$\alpha$. The observed Fe K$\alpha$ emission line is broadened. 

During the decay phase, we detected a narrow emission feature below 7\,keV. This feature plays a major role in shaping the whole iron-line region of the spectrum. If this narrow iron line is produced by the accretion disk and models the region with a single relativistic reflection model, the inferred inclination angle of the system is smaller than ${30^\circ}$ (see, e.g., Table 4 in \citealt{Shreeram_2020}), which contradicts known constraints for the inclination of GRS~1915+105 to be in the order of $i \simeq 60^\circ$~\citep[e.g.,][]{Reid_2014, Fender_1999}. When the inclination angle in our model is constrained to $60^\circ$, the iron-line region, in particular, the narrow component, cannot be well modeled. Therefore, we conclude that an additional component, a distant reflector, is necessary, and the observed complexity of the iron-line region is an interplay of the broad relativistic and narrow distant components. 

In addition to the narrow iron line, one unsolved feature remains in the residual plots: a suspected additional Si edge at $\sim$ 1.84\,keV. This additional Si edge was discussed, for example, by \citet{Lee_2002}, \citet{Martocchia_2006}, and \citet{Koljonen_2021}, and it can likely be attributed to calibration issues. To consider this additional absorption edge, we added an \texttt{edge} model whose threshold energy was fixed at 1.84\,keV, which we multiplied throughout the entire model. 

We thus had a simple phenomenological model for GRS~1915+105 in the soft and hard states. However, more interesting information on the accretion system can be extracted from the spectra with an improved physical model. We introduce an improved model for spectral fitting in the next section, where we use the NICER and HXMT data to constrain the broadband shape of the continuum. 

\subsubsection{An improved physical model for a simultaneous fit}\label{Sect:spec-timing:spectral:physical}
We started the joint fitting with the NICER and HXMT data to construct an improved physical model for the soft state before 2018 and the dimming phase after 2018 of GRS~1915+105. The energy coverage is broader when both instruments are used, and some key parameters such as the photon index $\Gamma$ and the intensity of the reflection component can be better constrained by the joint modeling. 

Based on the results provided by our preliminary fitting in Sect.~\ref{Sect:spec-timing:spectral:pre}, we used the Comptonization continuum multiplied by the neutral absorption model \texttt{TBabs}. We considered an additional Si absorption edge and a nonrelativistic reflection component as our initial improvement. We chose \texttt{xillverCp} to characterize this distant reflector~\citep{Garcia_2013}. 

With the distant reflector, we still had to consider the effect of the accretion disk. We used the fully relativistic reflection model \texttt{relxillCp} (from the \texttt{relxill} v2.3 package) to model the reflection produced on the accretion disk~\citep{Garcia_2014, Dauser_2014}. The indices of the emissivity profile $q_\text{in}$ and $q_\text{out}$ were fixed at 3, assuming a flat spacetime. The BH spin $a_*$ was fixed at 0.98~\citep{Miller_2013, Reid_2014} and the inclination angle was fixed at $60^\circ$~\citep{Reid_2014, Fender_1999}. We froze $R_\text{in}$ at the innermost stable circular orbit (ISCO). The outer radius of the disk was fixed at 400\,$r_g$, which can be considered sufficiently large. The indices of the power law $\Gamma$ and the temperatures of the electrons $kT_\text{e}$ in \texttt{relxillCp} and \texttt{xillverCp} were linked together with the parameters in the \texttt{Nthcomp} model because we assumed that the incident spectrum for the two reflectors is the same. The elemental abundance of the iron $A_\text{Fe}$ in the two reflection models were tied, and we first allowed it to vary between 0.5 and 2 in units of solar abundance. The ionization degrees $\log \xi$ and the electron densities $\log N$ in \texttt{relxillCp} and \texttt{xillverCp} were left free because we cannot assume the properties of the reflectors at this stage. We set \texttt{refl\_frac=-1} in the two reflection models, so that only the reflection fraction of the \texttt{relxillCp} and \texttt{xillverCp} was taken into account. Since both \texttt{relxillCp} and \texttt{xillverCp} were calculated with the inner temperature of the seed photons produced by the disk fixed at 0.01\,keV, we used a multiplicative table model \texttt{nthratio}\footnote{\url{https://github.com/garciafederico/nthratio}} to correct their slope in the low-energy band. 

We selected the table model \texttt{zxipcf} to characterize the ionization absorption lines, and it provided a rough estimate of the ionization degree~\citep{Reeves_2008}. We note that \texttt{zxipcf} was calculated based on \texttt{XSTAR}~\citep{Kallman_2004}, whose elemental abundances were given by \citet{Grevesse_1996} and are different from \citet{Wilms_2000}. In addition, the absorption features were calculated with a given ionization degree $\log \xi$ with the assumption that the source has an initial power-law spectrum of $\Gamma = 2$. This means that when the spectrum strongly deviates from a $\Gamma = 2$ power law, the estimated ionization parameter is unreliable, especially when the ionization parameter is required for further physical calculations. Nevertheless, considering all these caveats, this model was still our best choice because it is fast and universal for absorption from cold to hot plasma. We fixed the covering fraction of \texttt{zxipcf} at one and froze the redshift factor at zero, as the wind velocity cannot be constrained well due to the gain calibration of NICER. 

In \texttt{Xspec} parlance, the full model we used in the following spectral analysis can be written as
\begin{align}\label{eq-model}
    \texttt{TBabs} \times  \texttt{edge} &\times \texttt{zxipcf} \times \left( \texttt{Nthcomp} + \texttt{diskbb} \right. \notag \\
     + {} & \left. \texttt{nthratio} \times \left( \texttt{relxillCp} + \texttt{xillverCp} \right) \right) \, . 
\end{align}
We note that the spectra show no strong ionized absorption features in 2018 when the source was in the decay phase. Therefore, we omitted the \texttt{zxipcf} component in Eq.~\ref{eq-model}. 

The NICER-only observations were previously analyzed by \citet{Koljonen_2021}. Our model was developed using the HMXT data to constrain the continuum above 10\,keV, and while it is similar to the one employed in \citet{Koljonen_2021}, it shows several differences. First, we opted for \texttt{TBabs} combined with an additional Si edge to minimize the impact of the absorption model on the slope of the continuum. We incorporated emission lines from various elements using \texttt{xillverCp} and accounted for the absorption lines with \texttt{zxipcf} to accurately estimate the ionization degree and the flux ratios of the different spectral components. Finally, we employed \texttt{nthratio} to adjust the slopes of the reflection components to ensure consistency with the incident spectrum. 

We show examples of the joint results that fit this optimized modeling in panels (d) and (e) in Fig.~\ref{f-res}. We obtained better fit statistics than with the preliminary model, and almost all features in the residuals vanished. The fits show that the iron abundance $A_\text{Fe}$ and the electron density $\log N$ in \texttt{xillverCp} cannot be constrained in most of our observations, and we therefore later fixed $A_\text{Fe}$ to the solar value and set $\log N$ to 15. The ionization degree of \texttt{xillverCp} was also fixed at the lowest value ($\log \xi = 0$) as a cold distant reflector was assumed. 

\subsubsection{NICER-only modeling}
The joint fit shows that the obtained temperature of the electrons $kT_\text{e}$ reaches the upper limit of the allowed range (150\,keV) in the decay phase. While in the soft state, the $kT_\text{e}$ cannot be constrained even when the HXMT data are included. We let the electron temperature free when we fit the data before 2018. As mentioned before, we omitted the ionized absorption \texttt{zxipcf} for the observations from MJD\,58200 to MJD\,58500. Therefore, we only considered the ionized absorption for the observations in the soft state (before 2018) and at the end of the decay phase (after 2019). The thermal emission from the disk, the \texttt{diskbb} component, was omitted during the decay phase because the spectral fits suggested that the thermal emission was not necessary. 

\begin{figure*}
    \centering
    \includegraphics[width = 0.99\textwidth]{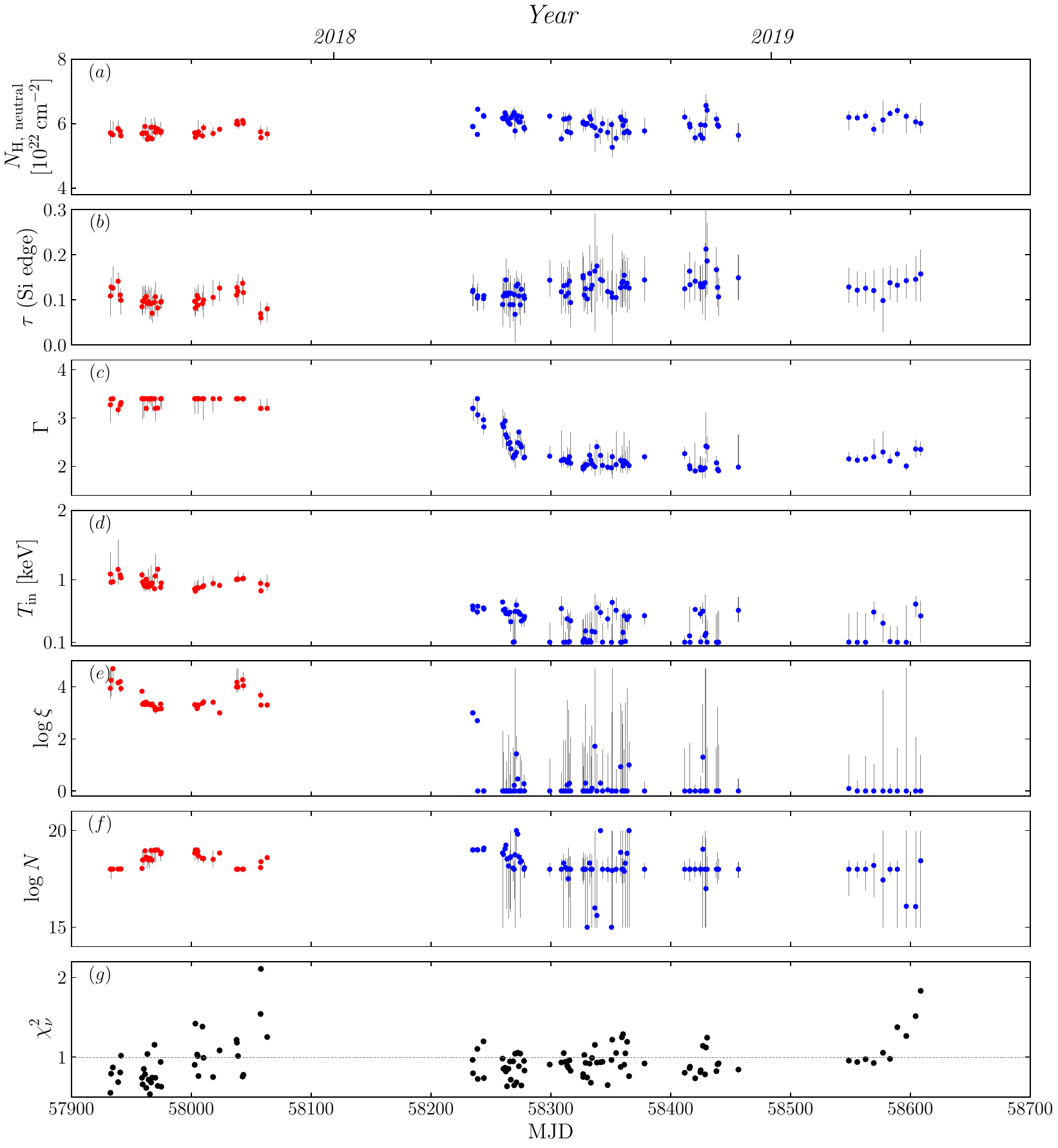}
    \caption{Evolution of the key parameters measured via the spectral analysis as a function of time. The panels show (from a to g) the equivalent hydrogen column $N_{\mathrm{H}}$ of the neutral absorption, the optical depth $\tau$ of the additional silicon edge, the photon index $\Gamma$, the temperature at the inner edge of the disk $T_{\mathrm{in}}$, the ionization degree $\log \xi$ of the accretion disk, the logarithmic number density of electrons of the accretion disk $\log N$, and the reduced chi-squares $\chi^2_\nu$ of the best-fits. The red and blue dots denote parameters obtained in the soft state and hard decay phase, respectively. }
    \label{f-specparams}
\end{figure*}

For most of our observations, no simultaneous NICER and HXMT data are available. We based the fits of the NICER-only data on the model developed above for the broader NICER + HXMT energy range with the following adaptations: We fixed the $kT_\text{e}$ at 150\,keV in the hard decay phase, but we let it free in the soft state. The iron abundance $A_\text{Fe}$ and the electron density $\log N$ of the distant reflector were fixed at the solar value and 15, respectively. The ionization degree $\log \xi$ of \texttt{xillverCp} was fixed at zero. The results of our fitting based on the NICER data are reported in Fig.~\ref{f-specparams}. 

\section{Results}\label{Sect:results}
\subsection{Before 2018}\label{Sect:results:soft}
GRS~1915+105 was in the soft state, where thermal emission dominated the spectrum. The column densities $N_{\mathrm{H}}$ of the absorption caused by the ISM is steadily measured at $(5.7 \pm 0.1) \times 10^{22}\, \text{cm}^{-2}$. The Comptonization exhibits photon indices approaching the upper limit of our model ($\Gamma \simeq 3.4$), and the inner temperatures $T_\text{in}$ of the accretion disk are approximately 1\,keV, which is higher than those during the decay phase (see panels $a$, $c$, and $d$ of Fig.~\ref{f-specparams}). The electron temperatures $kT_\text{e}$ are poorly constrained in the soft state, with a lower limit established at 5\,keV. The accretion disk is highly ionized, and the ionization degree $\log \xi$ is always greater than 3. 

Our estimated $T_\text{in}$, which is $\simeq 1\,$keV, is lower than previous studies (e.g., \citealt{Neilsen_2018} obtained a $T_\text{in}$ of 1.65--2.15\,keV based on the same NICER data). The reason likely is the lower estimate of the column density of the ISM absorption with our modeling. 

The ionized absorption feature is detected in all the NICER observations before 2018. The material is highly ionized with an ionization degree ($\log \xi$) greater than 4. The dips produced by the ionized absorption shown in the spectra are mostly the Fe {\sc xxvi} Ly$\alpha$ line at 7\,keV (see panel a of Fig.~\ref{f-res}). The column densities of the absorption are poorly constrained as the winds are highly ionized. 

\begin{figure*}
    \centering
    \includegraphics[width = 0.99\textwidth]{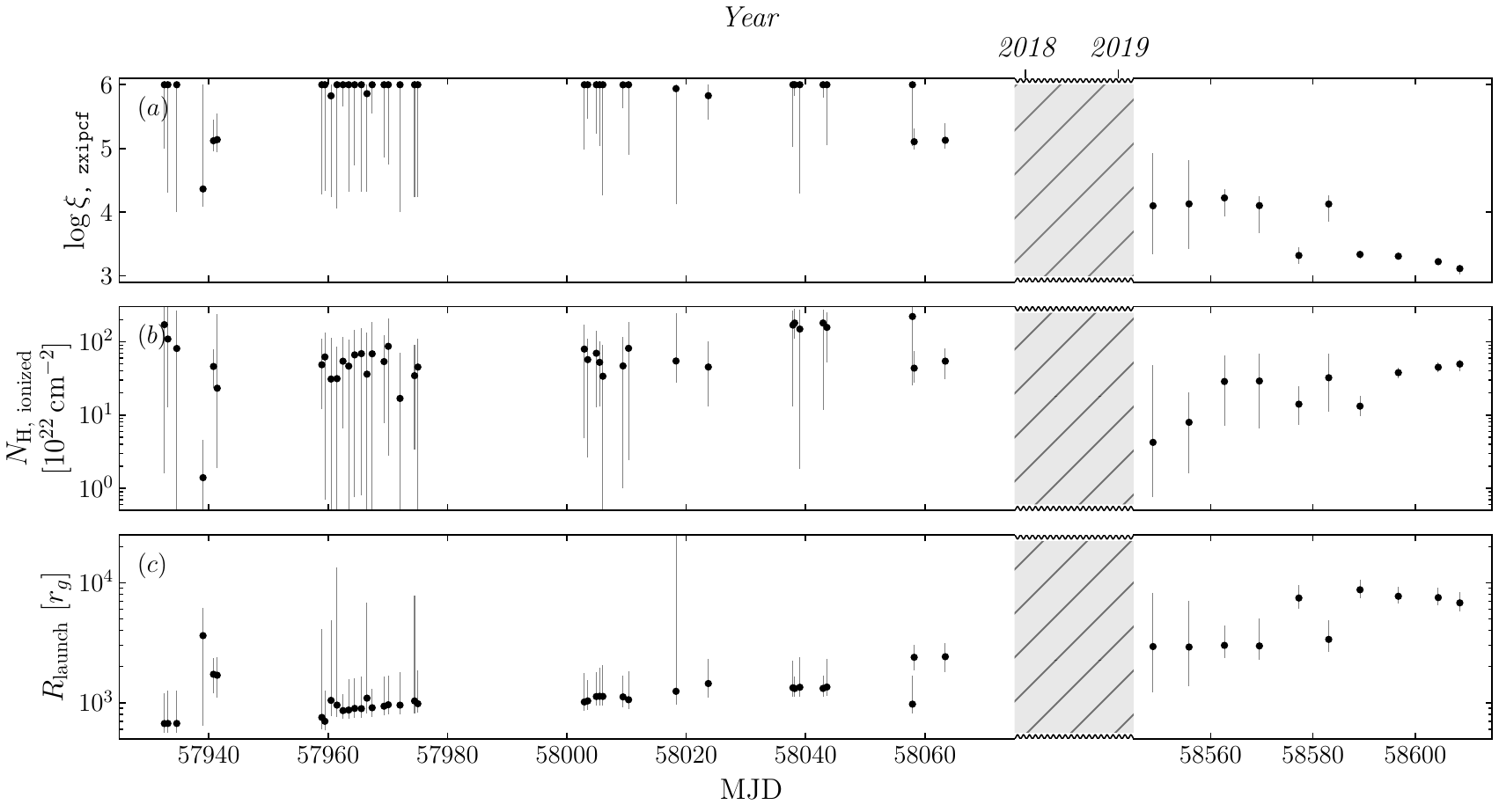}
    \caption{Evolution of the ionized wind absorption feature in the soft state and near the end of the decay phase. We omit the period from MJD\,58075 to 58545 as there is no strong evidence of an ionized absorption within this period. Panel ($a$): Ionization degree of the ionized absorption. Panel ($b$): Column density of the ionized absorption. Panel ($c$): Inferred launching radius of the ionized winds vs. time given a typical number density of the winds of $10^{14} \, \mathrm{cm}^{-3}$. }
    \label{f-ionized_abs}
\end{figure*}

We used the formula $R_\text{launch} \leq \sqrt{L/(N_\text{e} \cdot \xi)}$ to estimate the upper limit of the launching radius of the winds, where $L$ is the unabsorbed luminosity of the source~\citep{Tarter_1969}. The unabsorbed luminosity of the source continued at $\sim (18 \pm 8)\%\, L_\text{Edd}$ in the soft state before 2018. The electron number density $N_\text{e}$ of the winds is typically quoted as $10^{14} \, \mathrm{cm}^{-3}$ (following the method described in Sect. 4 of \citealt{Neilsen_2020}). The estimated launching radius with the fixed electron density is $10^3$--$10^4\,r_g$ and shows a significant negative correlation with the ionization degree. The results are shown in Fig.~\ref{f-ionized_abs}. No strong correlation between $R_\text{launch}$ and the disk inner temperature $T_\text{in}$ is observed. 

\subsection{During the decay phase}\label{Sect:results:hard}
With the modeling described in Sect.~\ref{Sect:spec-timing:spectral}, we obtained results from the spectral analysis by focusing on the exponential and linear decay phase. The behaviors of the source can be traced via the evolution of the key parameters. 

\subsubsection{The QPO-$\Gamma$ correlation}\label{Sect:results:hard:QPO}
We have illustrated the correlation between QPO-$\Gamma$ and a preliminary model in Fig.~\ref{f-qpo-freq}. In the same figure, we also presented their correlation with the improved physical model (Eq.~\ref{eq-model}) shown in black. The uncertainties of the newly estimated $\Gamma$ are larger as the model becomes more complex and has more degrees of freedom. We note that the new $\Gamma$ values are consistently higher than those measured with the preliminary model because the prominent iron-line feature is absorbed in a continuum produced by the reflection process of an incident Comptonization spectrum, which causes the slope of the primary emission to become steeper. The quasi-linear correlation between the main frequency of the QPOs and the photon index is still maintained. 

The evolution of the LFQPO central frequencies in Fig.~\ref{f-qpo-mjd} suggests that GRS~1915+105 begins in an intermediate state, transitions to a harder state after MJD\,58200, and returns to a hard-intermediate state at the end of the decay phase. The photon index $\Gamma$ of the Comptonization is mostly consistent with such a canonical state transition behavior. Although the strongly ionized winds are detected at the end of the decay phase, the photon index values increase simultaneously with the frequencies of the QPOs. This phenomenon indicates that the corona behaves independently from the ionized disk winds. Instead, the QPO frequencies are strongly related to the corona. 

The general quasi-linear correlation between the main QPO frequency and the photon index $\Gamma$ of the power-law component is commonly seen in most BHBs, for instance, Cyg~X-1, GX~339$-$4, and 4U~1630$-$47~\citep[see e.g.,][]{Tomsick_2000, Zdziarski_2004, Motta_2011, Boeck_2011, Grinberg_2014, Zhou_2022}. Although theories have been proposed to physically explain the origin of QPOs, for example, the Lense-Thirring precession~\citep{Bardeen_1975, Ingram_2009, Ingram_2011} or the accretion-ejection instability~\citep{Tagger_1999}, many of which assumed a truncated disk with hot flows inside, we cannot verify these theories by our results owing to the fixed inner radius of the accretion disk at ISCO in our model.

\subsubsection{The ionization degree and the electron density of the accretion disk}\label{Sect:results:hard:logxi}
Generally speaking, the influence of the ionization degree acting on the spectrum is complex. Different types of ions can produce large numbers of emission lines depending on the abundance of ions. These emission lines are broadened by the general relativistic effects, and the reflection spectrum is partly smoothed in the soft X-ray band. However, the iron emission lines at 6--7\,keV are usually not strongly affected by other emission lines because iron is more abundant than other heavier elements. The ionization degree can be precisely measured with a clear cutoff energy of the iron-line shape and a fixed inclination angle. 

\begin{figure}
    \centering
    \includegraphics[width = 0.49\textwidth]{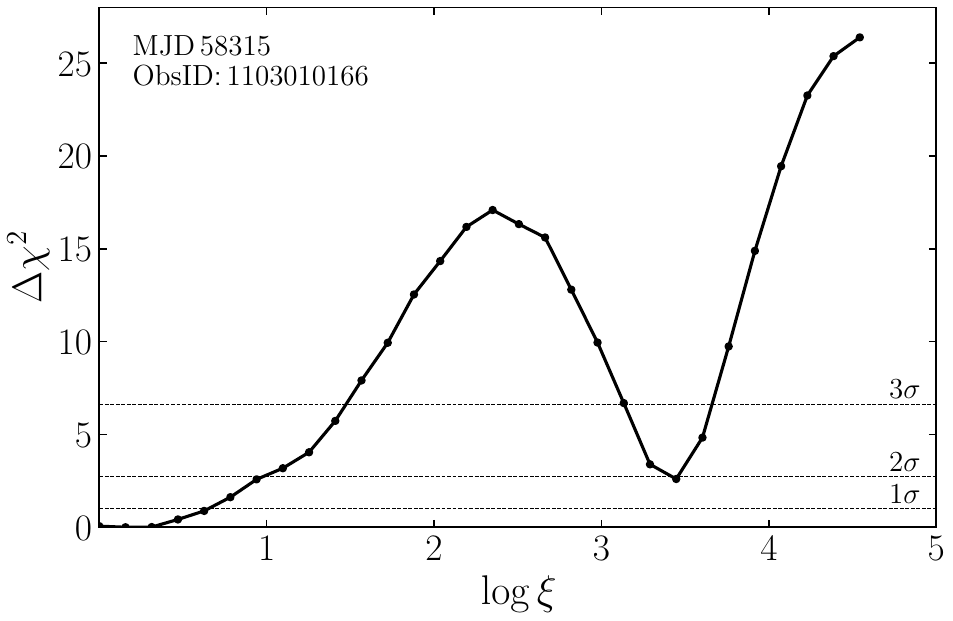}
    \caption{$\Delta \chi^2$ fitting statistics of an example spectrum in the decay phase, plotted vs. different values of the ionization degree $\log \xi$ of the relativistic reflection component \texttt{relxillCp}. The sample points are shown as black dots and are connected by solid lines. There are two local minima in this plot, with distinct ionization degrees. We discarded the solution with higher $\log \xi$ because it resulted in a vanishing primary emission originating from the corona. }
    \label{f-steppar}
\end{figure}

At the beginning of the decay phase, the ionization degree ($\log \xi$) is in the range of 2.7--3.0, close to the values we observed in the soft state. However, shortly after the start of the decay phase (since MJD\,58238), two branches of potential solutions with distinct ionization degrees and comparable statistical significance based on the spectral analysis develop. One example is shown in Fig.~\ref{f-steppar}. One possible solution branch has very low ionization degrees ($\log \xi \leq 2$), with the relative intensities of the relativistic reflection component divided by those of the primary emission from the corona being higher. The other branch typically exhibits a high ionization degree ($3 \leq \log \xi \leq 4$), but the fluxes from the primary emission tend to vanish in most of the spectra, causing the reflection fraction to approach an extremely high value, significantly greater than 10. Neither branch is universally superior to the other in a statistical sense. The strong degeneracy between the ionization degree and the relative intensity of spectral components arises from the asymptotic convergence between the reflection spectrum and the incident spectrum at high ionization degrees, where the material becomes transparent to X-ray photons. Statistically, both branches should be retained. However, the strong LFQPO features observed in the time domain indicate that the primary emission from the corona should not be too weak. Therefore, we decided to discard the branch of solutions with vanishing primary emission. 

Since MJD\,58238, the ionization degree ($\log \xi$) of the relativistic reflection component is constrained at low values for the majority of the observations (see panel $e$ in Fig.~\ref{f-specparams}), suggesting that the iron line is more likely located at 6.4\,keV in the rest frame of the accretion disk (the iron ions are in the form of Fe {\sc i} -- Fe {\sc xvii}; see \citealt{House_1969}). This result differs from the estimation in \citet{Koljonen_2021}, who obtained relatively high ionization degrees ($\log \xi \geq 3.5$). 

The logarithmic electron densities $\log N$ of the accretion disk cannot be well constrained for most of the observations during the decay phase. We obtained a mean value of $\log N \simeq 18$ from the best fits (see panel f of Fig.~\ref{f-specparams}). 

\subsubsection{The intensity of the reflection components}\label{Sect:results:hard:refl}
\begin{figure}
    \centering
    \includegraphics[width = 0.49\textwidth]{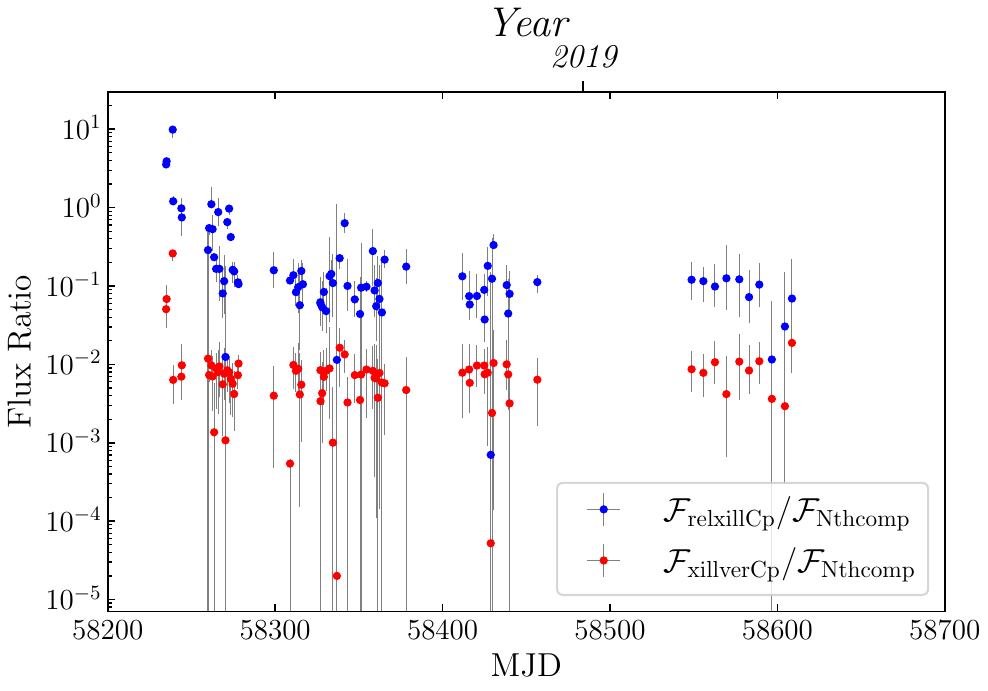}
    \caption{Evolution of the ratio of the unabsorbed flux of the incident Comptonization spectra and that of the relativistic and distant reflection components vs. time. The fluxes of \texttt{relxillCp}, which models the relativistic reflector, and those of \texttt{xillverCp}, which models the distant reflector, were corrected by the multiplicative model \texttt{nthratio} (see Sect.~\ref{Sect:spec-timing:spectral:physical}). $\mathcal{F}_\text{relxillCp}/\mathcal{F}_\text{Nthcomp}$ indicates the geometrical information of the accreting system. }
    \label{f-fluxratio-mjd}
\end{figure}

Assuming that the emission from the corona is dominant and discarding the branch of solutions with higher $\log \xi$ values, we computed the intensity of each spectral component in our modeling in Eq.~\ref{eq-model}. The flux ratios of the (non-)relativistic reflection components and the incident primary emission with its 1$\sigma$ uncertainties were computed, and the results are shown in Fig.~\ref{f-fluxratio-mjd}. 

We first find that the flux ratio  $\mathcal{F}_\text{xillverCp}/\mathcal{F}_\text{Nthcomp}$ drops rapidly within the first ten days of the decay phase (MJD\,58234--58244) and then steadily remains weaker by two orders of magnitude than the primary emission for the remaining time. When we assume that the distant reflector remains stable over a ten-day period, the sudden drop in the flux ratio $\mathcal{F}_\text{xillverCp}/\mathcal{F}_\text{Nthcomp}$ is likely due to the self-obscuration of the source, that is, the radiation from the accretion center was impeded by the thickness of the inner accretion disk. In contrast, the flux ratio $\mathcal{F}_\text{relxillCp}/\mathcal{F}_\text{Nthcomp}$ continues to decrease after the self-obscuration period, and this decline lasted until MJD\,58280 and finally reached approximately one-tenth of the primary emission. It maintained this relative strength until the end of the decay phase. 

\subsubsection{The ionized absorption near the end of the decay phase (MJD\,58548--58608)}
The emission generated by GRS~1915+105 underwent a warmly ionized wind absorption near the end of the linear decay~\citep{Miller_2020, Koljonen_2021}. With the quasi-weekly NICER observations between MJD\,58548 and MJD\,58608, we measured the column densities and the ionization degrees of the wind absorption. The results with their 2$\sigma$ uncertainties are presented in the right part of Fig.~\ref{f-ionized_abs}. 

We observe a gradually strengthening absorption column with a slowly decreasing ionization degree $\log \xi$ from $\sim$ 4 to 3 within 60 days. In the last few observations, the complex features between 6.4\,keV and 6.7\,keV produced by the warm ionized absorber are detected. Highly ionized winds are usually observed in the soft state without QPO features in the time domain~\citep{Ponti_2012}. However, GRS~1915+105 exhibited strong evidence of a hard-intermediate state at the end of the decay phase, with intermediate $\Gamma$ values ($2.0 \leq \Gamma \leq 2.4$) and clear type-C QPO signals. 

Similar to Sect.~\ref{Sect:results:soft}, the launching radius of the wind was calculated with a fixed electron density of $10^{14} \, \mathrm{cm}^{-3}$. We thus observe a slightly increasing launching radius with relatively large uncertainties during the end of the decay phase (see panel $c$ of Fig.~\ref{f-ionized_abs}) with this fixed electron density. However, we note that the number density of the wind is highly dependent on the radius and polar angle of the disk ~\citep{Fukumura_2017}. Thus, the estimation of the launching radius of the wind here is a qualitative response to the ionization degree, as the unabsorbed luminosity of the source maintains $\sim 2\%\,L_\text{Edd}$ during the end of the decay phase. 

We note that although the model \texttt{zxipcf} characterizing the ionized absorption was computed with an incident power-law spectrum of $\Gamma = 2$, which may cause a deviation when estimating the ionization degree as the incident spectrum is not a pure power-law with $\Gamma = 2$, the estimated ionization degree $\log \xi$ is not expected to deviate much from the physical value in our study. The uncertainties of $\log \xi$ absorb the systematic errors made by the incident spectrum due to the highly ionized plasma outflowing in the soft state, and the $\Gamma$ values do not strongly deviate from 2 at the end of the decay phase. 

\section{Discussion}\label{Sect:discussion}
\subsection{A rarely seen warm ionized wind in the hard-intermediate state}\label{Sect:discussion:wind}
Since the discovery of the disk winds, the majority of the confirmed ionized wind cases detected in the iron band were found in the soft state of the BHBs and with a high ionization degree ($\log \xi \geq 4$)~\citep[see e.g.,][]{Miller_2006_Nature, Miller_2006_ApJ, Kubota_2007, Neilsen_2009, Ponti_2012}. Although there are several reports on the ionized absorption observed in the hard or hard-intermediate state, some of the detections depend on the choice of the reflection models~\citep{Xu_2018, Wang_2021}. \citet{Lee_2002} reported a highly ionized wind ($\log \xi \simeq 4.15$) in GRS~1915+105 when the source was in the hard $\chi$ state, but the unabsorbed luminosity of the source reached $\sim 40 \% \, L_\text{Edd}$ at that time. It is natural to produce a highly ionized wind with such a high luminosity. \citet{Shidatsu_2013} detected a warmly ionized wind in the hard state of MAXI~J1305$-$704, with an ionization degree of $\log \xi \sim 2$. 

The behaviors of the absorber are dependent on the assumptions. With a variable electron density by assuming that the thickness of the wind $\Delta R$ is comparable to the launching radius $R_\text{launch}$, the electron density $N_\text{e}$ can be estimated by $N_\text{e} = N_\text{H}/\Delta R \sim N_\text{H}/R_\text{launch}$. Therefore, $R_\text{launch} \sim L/(N_\text{H} \cdot \xi)$. The mean value of the estimated $N_\text{e}$ at the end of the decay is about $10^{13} \, \mathrm{cm}^{-3}$, which does not result in a change in $R_\text{launch}$ by more than one order of magnitude. With a constant $N_\text{e}$ of $10^{14} \, \mathrm{cm}^{-3}$, the estimated $R_\text{launch}$ of the absorber instead shows a potentially extremely slow outward movement with an upper-limit speed of $\dot{R}_\text{launch} \simeq 0.02 \, \mathrm{km} \cdot \mathrm{s}^{-1}$. Compared with the velocity of the wind itself ($v \geq {10}^2 \, \mathrm{km} \cdot \mathrm{s}^{-1}$; see \citealt{Miller_2020}), the motion of the launching radius is negligible. 

The estimated launching radii of the winds in GRS~1915+105 during the soft state before 2018 and at the end of the decay phase in 2019 are both in the range of $10^3$--$10^4\,r_g$, which suggests that the launching mechanism of the outflow likely was the same. Although the obscuration of this source after May 2019 was attributed to local absorption~\citep{Neilsen_2020, Balakrishnan_2021}, the cause of the decay before the obscuration is different. The neutral absorption $N_\mathrm{H,\ neutral}$ maintains $6 \times 10^{22}\,\mathrm{cm^{-2}}$ throughout the decay and shows no significant enhancement until the onset of the obscured state. The cold matter seen in the obscured state likely originates from the outflow that formed near the end of the decay. 

\subsection{The interaction between the corona and the accretion disk}
In Sect.~\ref{Sect:results:hard:refl} we presented the intensity analysis of the primary emission and the (non-)relativistic reflection components (see Fig.~\ref{f-fluxratio-mjd}). The ratio $\mathcal{F}_\text{relxillCp}/\mathcal{F}_\text{Nthcomp}$ and $\mathcal{F}_\text{xillverCp}/\mathcal{F}_\text{Nthcomp}$ may unveil the geometrical information of the accreting system. 

 $\mathcal{F}_\text{relxillCp}/\mathcal{F}_\text{Nthcomp}$ and $\mathcal{F}_\text{xillverCp}/\mathcal{F}_\text{Nthcomp}$ dropped to one-tenth of their original values during the first 10 days of the decay phase (MJD\,58234--58244). These drastic decreases were likely produced by the fading of the self-obscuration of the accreting system. $\mathcal{F}_\text{xillverCp}/\mathcal{F}_\text{Nthcomp}$ remained at $10^{-2}$ after the 10-day period, but $\mathcal{F}_\text{relxillCp}/\mathcal{F}_\text{Nthcomp}$ continued to decrease from $10^{0}$ to $10^{-1}$ in the following 30 days (MJD\,58245--58275). The decrease observed solely in $\mathcal{F}_\text{relxillCp}/\mathcal{F}_\text{Nthcomp}$ was likely associated with the geometrical changes in the accreting system during an intermediate-to-hard transition. 

The reflection component in BHBs becomes stronger in general when the spectrum is softer because the Comptonization in soft states is weaker~\citep{Steiner_2016}. Therefore, the ratio $\mathcal{F}_\text{relxillCp}/\mathcal{F}_\text{Nthcomp}$ is assumed to become smaller in the intermediate-to-hard transition and vice versa. Although the states exhibited by GRS~1915+105 are generally not strictly the canonical hard or soft states during an outburst, it can still undergo transitions that are considered part of the canonical state evolution due to the long-term instability of the corona. Similar incomplete transitions have also been found in other persistent sources, for example, in Cyg~X-1~\citep{Zhou_2022}. The reduced flux ratio at the beginning of the decay phase aligns with this expectation. We expect to observe an increase in this ratio at the end of the decay phase if the latter transition is normal because the photon indices $\Gamma$ and the frequencies of the QPO increase then. However, the flux ratio remained low and even exhibited a decreasing trend until the end of the decay phase, which suggests that the transition near the end of the decay phase is different from a normal softening state transition. 

In the context of the lamp-post model, in which the corona is simplified as a point-like source above the disk, the two ratios can easily be interpreted as signals of the distances between the corona and two different reflectors. An increasing ratio means a closer distance, and vice versa. The evolution of $\mathcal{F}_\text{relxillCp}/\mathcal{F}_\text{Nthcomp}$ suggests that the corona consistently recedes from the accretion disk in this scenario. However, recent polarization results indicate that the primary photon emitter could be parallel to the accretion disk rather than a source perpendicular to the accretion disk (\citealt{Krawczynski_2022}; \citealt{Veledina_2023}; however, see e.g., \citealt{Dexter_2024} for a different interpretation). In this scenario, the ratio $\mathcal{F}_\text{relxillCp}/\mathcal{F}_\text{Nthcomp}$ works as an indicator of the strength of the interaction between the corona and the standard accretion disk. In the normal hard state, the optical depth of the corona is higher, which causes more photons to be scattered and dilutes the initial relativistic effects of the disk photons~\citep{Petrucci_2001}. In contrast, the source exhibits a weak interaction at the end of the decay phase. This odd phenomenon might be attributed to the changed geometry of the accretion system, for instance, a relatively shrinking standard disk due to the lack of infalling material. Returning radiation can also play an important role in producing reflection components~\citep{Mirzaev_2024}. The photons emitted from the inner part of the disk have a higher probability of arriving at the surface of the accretion disk and re-irradiating the disk again~\citep{Dauser_2022}. An increasing inner radius of the disk during this period can weaken the irradiation from the returning radiation and thus reduce the flux of the relativistic reflection component. 

\subsection{The interplay between the disk winds and the corona during and after the decay}\label{Sect:discussion:interplay}
The physical mechanisms in the accretion process of BHBs are still debated, but we benefit from observational evidence to gradually disentangle the puzzle of accretion. The jets and disk winds usually exhibit mutually exclusive interplay within a normal outburst~\citep{Neilsen_2009, Ponti_2012}, although exceptions exist~\citep{Miller_2006_ApJ, Kalemci_2016, Homan_2016, Motta_2021}. The behaviors of the jet are considered to be closely related to the corona~\citep[see e.g.,][]{Markoff_2005}. 

In GRS~1915+105, we have confirmed that the source underwent a partially hardening-to-softening transition during the decay phase, where the ionized wind absorption features merely appeared near the end of the decay. Type-C QPOs were detected throughout the decay phase, and their frequencies were strongly correlated to the photon indices of the Comptonization component. The spectral-timing behaviors of the corona seem to be independent from the emergence of the ionized disk outflow. 

When we assume that the disk winds that were observed in the soft state and the hard decay phase share the same launching mechanism, irreconcilable contradictions will arise when the appearance of the disk wind is forcibly linked with the vanishing of the jet or the softening of the spectrum. The radio emission and the disk wind clearly coexist at the end of the decay phase~\citep{Motta_2021}. It is not very feasible either to connect the appearance of disk winds with the movement of $r_\text{in}$, the inner radius of the disk, as $r_\text{in}$ approaches the ISCO in the soft state but increases in the scenario of returning radiation. However, considering that the launching position of the wind is far from the binary center, we suggest a potential hysteretic interaction between the corona and the periphery of the accretion disk. Typically, suppressed jets release less energy into the environment, and more energy is directed toward Compton heating, which accelerates the winds~\citep{Narayan_1995, Merloni_2002}. These winds carry angular momentum away, which slows the mass inflow down and promotes faster thermal equilibrium in the inner regions of the accretion disk~\citep{Blandford_1999}. However, during the decay phase of GRS 1915+105, winds driven by a relatively low luminosity would further decelerate the infalling matter, which would lead to a more rapid fading of the accretion system and eventually to the emergence of the obscured state. 

\citet{Neilsen_2020} reported a fast flare observed in the obscured state of GRS~1915+105. Fe {\sc xxvi} Ly$\alpha$ and Fe {\sc xxv} He$\alpha$ lines were detected in emission before the flare and in absorption when the count rate increased, indicating that the highly ionized outflow always existed during the focused observation: Absorption lines are visible when a strong central source back-lights the medium, and emission lines emerge when the central source becomes faint. In periods with a lower count rate, three prominent emission lines (quasi-neutral Fe K$\alpha$ line at 6.4\,keV, Fe {\sc xxv} He$\alpha$ line at 6.7\,keV, and Fe {\sc xxvi} Ly$\alpha$ at 7\,keV) were observed~\citep{Miller_2020, Koljonen_2021} in the Fe band. The origins of the three lines are different, as the quasi-neutral Fe lines come from the cooling area of the accretion disk and the ionized Fe lines come from the highly ionized plasma that flows out of the disk. The emergence of three Fe emission lines is not unique to GRS~1915+105 and was found in other systems such as V404~Cygni~\citep{King_2015} and V4641~Sgr~\citep{Shaw_2022}, which are well-known for their high inclination and variable local obscuration~\citep{Koljonen_2020}. 

\section{Conclusions}\label{Sect:conclusion}
We have conducted a spectral-timing analysis of GRS~1915+105, for which we used all available NICER and Insight-HXMT data before MJD\,58617, when the source entered the obscured state. In particular, we used simultaneous NICER and HXMT data to develop a model that described the broadband spectrum of the source, and we used this model with the higher-cadence NICER data to trace the evolution of the source. 

In the disk-dominated state before 2018, we consistently observed a highly ionized wind with the detection of the Fe {\sc xxvi} Ly$\alpha$ line at 7\,keV. The photon index approached the upper limit of our model ($\Gamma \simeq 3.4$), and the intensity of the thermal radiation dominated the spectrum. 

During the decay phase, which corresponds to a spectral transition from a partially hardening process to an unusual obscuring process, we found a quasi-linear correlation between the photon index $\Gamma$ and the frequency of the main QPO with an empirical model and an improved physical model. At the end of the decay phase, the spectrum showed strong absorption features with a decreasing ionization degree when the source was in the hard-intermediate state, with clear type-C QPO features. The estimated launching radius of this warmly ionized outflow appears to be on the same order of magnitude as that in the soft state. This indicates that the launching mechanism of the winds is likely the same in the soft and hard states near the end of the decay. 

At the beginning of the decay phase, the relative intensity of the relativistic and nonrelativistic reflection components was strong, but it weakened by a factor of 10 within 10 days, possibly due to the vanishing self-obscuration in the accreting system. The relativistic reflection component continued to weaken over the next 30 days, while the nonrelativistic reflection component remained steady. The two reflection components maintained their relative strengths in the middle of the decay phase. However, the relativistic component even showed a potentially decreasing trend as warmly ionized winds appeared near the end of the decay. This discrepancy corresponds to an unusual softening process that either indicates an increasing distance between the disk and the corona (in the context of the lamp-post model) or a rising inner radius of the disk (in the context of returning radiation). The QPO frequencies and photon index are strongly interdependent, and the launch of ionized winds further slowed the accreting of the infalling matter down. 

\begin{acknowledgements}
M.~Z. would like to thank the support from the China Scholarship Council (CSC 202006100027). This research has made use of NASA’s Astrophysics Data System Bibliographic Services. This research also made use of \texttt{ISIS} functions (\texttt{isisscripts}\footnote{\url{http://www.sternwarte.uni-erlangen.de/isis/}}) provided by ECAP/Remeis observatory and MIT. This work made use of data from the Insight--HXMT mission, a project funded by the China National Space Administration (CNSA) and the Chinese Academy of Sciences (CAS). L. D. Kong is grateful for the financial support provided by the Sino-German (CSC-DAAD) Postdoc Scholarship Program (57607866).
\end{acknowledgements}

\bibliographystyle{aa}
% \bibliography{aa_abbrv, mnemonic, references}
\bibliography{references}

\begin{thebibliography}{114}
\expandafter\ifx\csname natexlab\endcsname\relax\def\natexlab#1{#1}\fi

\bibitem[{{Arnaud}(1996)}]{Arnaud_1996}
{Arnaud}, K.~A. 1996, in Astronomical Society of the Pacific Conference Series, Vol. 101, Astronomical Data Analysis Software and Systems V, ed. G.~H. {Jacoby} \& J.~{Barnes}, 17

\bibitem[{{Athulya} {et~al.}(2022){Athulya}, {Radhika}, {Agrawal}, {Ravishankar}, {Naik}, {Mandal}, \& {Nandi}}]{Athulya_2022}
{Athulya}, M.~P., {Radhika}, D., {Agrawal}, V.~K., {et~al.} 2022, \mnras, 510, 3019

\bibitem[{{Bachetti} {et~al.}(2021){Bachetti}, {Huppenkothen}, {Khan}, {Mishra}, {Sharma}, {Stevens}, {Desai}, {Rashid}, {Martinez Ribeiro}, {Swinbank}, {Sip{\H{o}}cz}, {tappina}, {omargamal8}, {Davis}, {Rasquinha}, {Balm}, {Mumford}, {Campana}, {Garg}, {Tandon}, {Hota}, {Nick}, {Raj}, {Mishra}, {Smith}, {Mahlke}, {Sachidanand}, {Kumar}, {Vall{\'e}s Blanco}, \& {Kothari}}]{Bachetti_2021}
{Bachetti}, M., {Huppenkothen}, D., {Khan}, U., {et~al.} 2021, {StingraySoftware/stingray: Version 0.3}

\bibitem[{{Balakrishnan} {et~al.}(2021){Balakrishnan}, {Miller}, {Reynolds}, {Kammoun}, {Zoghbi}, \& {Tetarenko}}]{Balakrishnan_2021}
{Balakrishnan}, M., {Miller}, J.~M., {Reynolds}, M.~T., {et~al.} 2021, \apj, 909, 41

\bibitem[{{Bardeen} \& {Petterson}(1975)}]{Bardeen_1975}
{Bardeen}, J.~M. \& {Petterson}, J.~A. 1975, \apjl, 195, L65

\bibitem[{{Begelman} {et~al.}(1983){Begelman}, {McKee}, \& {Shields}}]{Begelman_1983}
{Begelman}, M.~C., {McKee}, C.~F., \& {Shields}, G.~A. 1983, \apj, 271, 70

\bibitem[{{Belloni} {et~al.}(2000){Belloni}, {Klein-Wolt}, {M{\'e}ndez}, {van der Klis}, \& {van Paradijs}}]{Belloni_2000}
{Belloni}, T., {Klein-Wolt}, M., {M{\'e}ndez}, M., {van der Klis}, M., \& {van Paradijs}, J. 2000, \aap, 355, 271

\bibitem[{{Blandford} \& {Begelman}(1999)}]{Blandford_1999}
{Blandford}, R.~D. \& {Begelman}, M.~C. 1999, \mnras, 303, L1

\bibitem[{{B{\"o}ck} {et~al.}(2011){B{\"o}ck}, {Grinberg}, {Pottschmidt}, {Hanke}, {Nowak}, {Markoff}, {Uttley}, {Rodriguez}, {Pooley}, {Suchy}, {Rothschild}, \& {Wilms}}]{Boeck_2011}
{B{\"o}ck}, M., {Grinberg}, V., {Pottschmidt}, K., {et~al.} 2011, \aap, 533, A8

\bibitem[{{Bogdanov} {et~al.}(2019){Bogdanov}, {Guillot}, {Ray}, {Wolff}, {Chakrabarty}, {Ho}, {Kerr}, {Lamb}, {Lommen}, {Ludlam}, {Milburn}, {Montano}, {Miller}, {Baub{\"o}ck}, {{\"O}zel}, {Psaltis}, {Remillard}, {Riley}, {Steiner}, {Strohmayer}, {Watts}, {Wood}, {Zeldes}, {Enoto}, {Okajima}, {Kellogg}, {Baker}, {Markwardt}, {Arzoumanian}, \& {Gendreau}}]{Bogdanov_2019}
{Bogdanov}, S., {Guillot}, S., {Ray}, P.~S., {et~al.} 2019, \apjl, 887, L25

\bibitem[{{Cao} {et~al.}(2020){Cao}, {Jiang}, {Meng}, {Zhang}, {Luo}, {Yang}, {Zhang}, {Gu}, {Sun}, {Liu}, {Yang}, {Li}, {Tan}, {Liu}, {Du}, {Lu}, {Xu}, {Guan}, {Zhang}, {Wang}, {Li}, {Zhang}, {Wen}, {Qu}, {Song}, {Li}, {Ge}, {Zhou}, {Xiong}, {Zhang}, {Zhang}, {Cheng}, {Zhang}, {Li}, {Liang}, {Gao}, {Yang}, {Liu}, {Liu}, {Yang}, \& {Zhang}}]{Cao_2020_me}
{Cao}, X., {Jiang}, W., {Meng}, B., {et~al.} 2020, Science China Physics, Mechanics, and Astronomy, 63, 249504

\bibitem[{{Castro-Tirado} {et~al.}(1992){Castro-Tirado}, {Brandt}, \& {Lund}}]{Castro-Tirado_1992}
{Castro-Tirado}, A.~J., {Brandt}, S., \& {Lund}, N. 1992, \iaucirc, 5590, 2

\bibitem[{{Castro-Tirado} {et~al.}(1994){Castro-Tirado}, {Brandt}, {Lund}, {Lapshov}, {Sunyaev}, {Shlyapnikov}, {Guziy}, \& {Pavlenko}}]{Castro-Tirado_1994}
{Castro-Tirado}, A.~J., {Brandt}, S., {Lund}, N., {et~al.} 1994, \apjs, 92, 469

\bibitem[{{Chen} {et~al.}(2020){Chen}, {Cui}, {Li}, {Wang}, {Xu}, {Lu}, {Wang}, {Chen}, {Han}, {Hu}, {Zhang}, {Huo}, {Yang}, {Li}, {Lu}, {Zhang}, {Li}, {Zhang}, {Xiong}, {Zhang}, {Xue}, {Zhao}, {Zhu}, {Zhu}, {Liu}, {Yang}, \& {Zhang}}]{Chen_2020_le}
{Chen}, Y., {Cui}, W., {Li}, W., {et~al.} 2020, Science China Physics, Mechanics, and Astronomy, 63, 249505

\bibitem[{{Dauser} {et~al.}(2014){Dauser}, {Garcia}, {Parker}, {Fabian}, \& {Wilms}}]{Dauser_2014}
{Dauser}, T., {Garcia}, J., {Parker}, M.~L., {Fabian}, A.~C., \& {Wilms}, J. 2014, \mnras, 444, L100

\bibitem[{{Dauser} {et~al.}(2022){Dauser}, {Garc{\'\i}a}, {Joyce}, {Licklederer}, {Connors}, {Ingram}, {Reynolds}, \& {Wilms}}]{Dauser_2022}
{Dauser}, T., {Garc{\'\i}a}, J.~A., {Joyce}, A., {et~al.} 2022, \mnras, 514, 3965

\bibitem[{{Dexter} \& {Begelman}(2024)}]{Dexter_2024}
{Dexter}, J. \& {Begelman}, M.~C. 2024, \mnras, 528, L157

\bibitem[{{Done} {et~al.}(2018){Done}, {Tomaru}, \& {Takahashi}}]{Done_2018}
{Done}, C., {Tomaru}, R., \& {Takahashi}, T. 2018, \mnras, 473, 838

\bibitem[{{Fender} {et~al.}(1999){Fender}, {Garrington}, {McKay}, {Muxlow}, {Pooley}, {Spencer}, {Stirling}, \& {Waltman}}]{Fender_1999}
{Fender}, R.~P., {Garrington}, S.~T., {McKay}, D.~J., {et~al.} 1999, \mnras, 304, 865

\bibitem[{{Fender} {et~al.}(2006){Fender}, {Stirling}, {Spencer}, {Brown}, {Pooley}, {Muxlow}, \& {Miller-Jones}}]{Fender_2006}
{Fender}, R.~P., {Stirling}, A.~M., {Spencer}, R.~E., {et~al.} 2006, \mnras, 369, 603

\bibitem[{{Fukumura} {et~al.}(2017){Fukumura}, {Kazanas}, {Shrader}, {Behar}, {Tombesi}, \& {Contopoulos}}]{Fukumura_2017}
{Fukumura}, K., {Kazanas}, D., {Shrader}, C., {et~al.} 2017, Nature Astronomy, 1, 0062

\bibitem[{{Garc{\'\i}a} {et~al.}(2014){Garc{\'\i}a}, {Dauser}, {Lohfink}, {Kallman}, {Steiner}, {McClintock}, {Brenneman}, {Wilms}, {Eikmann}, {Reynolds}, \& {Tombesi}}]{Garcia_2014}
{Garc{\'\i}a}, J., {Dauser}, T., {Lohfink}, A., {et~al.} 2014, \apj, 782, 76

\bibitem[{{Garc{\'\i}a} {et~al.}(2013){Garc{\'\i}a}, {Dauser}, {Reynolds}, {Kallman}, {McClintock}, {Wilms}, \& {Eikmann}}]{Garcia_2013}
{Garc{\'\i}a}, J., {Dauser}, T., {Reynolds}, C.~S., {et~al.} 2013, \apj, 768, 146

\bibitem[{{Gendreau} {et~al.}(2016){Gendreau}, {Arzoumanian}, {Adkins}, {Albert}, {Anders}, {Aylward}, {Baker}, {Balsamo}, {Bamford}, {Benegalrao}, {Berry}, {Bhalwani}, {Black}, {Blaurock}, {Bronke}, {Brown}, {Budinoff}, {Cantwell}, {Cazeau}, {Chen}, {Clement}, {Colangelo}, {Coleman}, {Coopersmith}, {Dehaven}, {Doty}, {Egan}, {Enoto}, {Fan}, {Ferro}, {Foster}, {Galassi}, {Gallo}, {Green}, {Grosh}, {Ha}, {Hasouneh}, {Heefner}, {Hestnes}, {Hoge}, {Jacobs}, {J{\o}rgensen}, {Kaiser}, {Kellogg}, {Kenyon}, {Koenecke}, {Kozon}, {LaMarr}, {Lambertson}, {Larson}, {Lentine}, {Lewis}, {Lilly}, {Liu}, {Malonis}, {Manthripragada}, {Markwardt}, {Matonak}, {Mcginnis}, {Miller}, {Mitchell}, {Mitchell}, {Mohammed}, {Monroe}, {Montt de Garcia}, {Mul{\'e}}, {Nagao}, {Ngo}, {Norris}, {Norwood}, {Novotka}, {Okajima}, {Olsen}, {Onyeachu}, {Orosco}, {Peterson}, {Pevear}, {Pham}, {Pollard}, {Pope}, {Powers}, {Powers}, {Price}, {Prigozhin}, {Ramirez}, {Reid}, {Remillard}, {Rogstad}, {Rosecrans}, {Rowe}, {Sager}, {Sanders},
  {Savadkin}, {Saylor}, {Schaeffer}, {Schweiss}, {Semper}, {Serlemitsos}, {Shackelford}, {Soong}, {Struebel}, {Vezie}, {Villasenor}, {Winternitz}, {Wofford}, {Wright}, {Yang}, \& {Yu}}]{Gendreau_2016}
{Gendreau}, K.~C., {Arzoumanian}, Z., {Adkins}, P.~W., {et~al.} 2016, in Society of Photo-Optical Instrumentation Engineers (SPIE) Conference Series, Vol. 9905, Space Telescopes and Instrumentation 2016: Ultraviolet to Gamma Ray, ed. J.-W.~A. {den Herder}, T.~{Takahashi}, \& M.~{Bautz}, 99051H

\bibitem[{Grevesse {et~al.}(1996)Grevesse, Noels, Sauval, Holt, \& Sonneborn}]{Grevesse_1996}
Grevesse, N., Noels, A., Sauval, A., Holt, S., \& Sonneborn, G. 1996, in ASP Conf. Ser, Vol.~99, 117

\bibitem[{{Grinberg} {et~al.}(2014){Grinberg}, {Pottschmidt}, {B{\"o}ck}, {Schmid}, {Nowak}, {Uttley}, {Tomsick}, {Rodriguez}, {Hell}, {Markowitz}, {Bodaghee}, {Cadolle Bel}, {Rothschild}, \& {Wilms}}]{Grinberg_2014}
{Grinberg}, V., {Pottschmidt}, K., {B{\"o}ck}, M., {et~al.} 2014, \aap, 565, A1

\bibitem[{{Guo} {et~al.}(2020){Guo}, {Liao}, {Zhang}, {Zhang}, {Tan}, {Song}, {Lu}, {Cao}, {Chang}, {Chen}, {Du}, {Ge}, {Gu}, {Jiang}, {Jin}, {Li}, {Li}, {Li}, {Liu}, {Liu}, {Lu}, {Luo}, {Meng}, {Sun}, {Yang}, {Yang}, {You}, {Zhang}, {Zhao}, \& {Zhang}}]{Guo_2020_mebackground}
{Guo}, C.-C., {Liao}, J.-Y., {Zhang}, S., {et~al.} 2020, Journal of High Energy Astrophysics, 27, 44

\bibitem[{{Hannikainen} {et~al.}(2005){Hannikainen}, {Rodriguez}, {Vilhu}, {Hjalmarsdotter}, {Zdziarski}, {Belloni}, {Poutanen}, {Wu}, {Shaw}, {Beckmann}, {Hunstead}, {Pooley}, {Westergaard}, {Mirabel}, {Hakala}, {Castro-Tirado}, \& {Durouchoux}}]{Hannikainen_2005}
{Hannikainen}, D.~C., {Rodriguez}, J., {Vilhu}, O., {et~al.} 2005, \aap, 435, 995

\bibitem[{{Higginbottom} \& {Proga}(2015)}]{Higginbottom_2015}
{Higginbottom}, N. \& {Proga}, D. 2015, \apj, 807, 107

\bibitem[{{Homan} {et~al.}(2016){Homan}, {Neilsen}, {Allen}, {Chakrabarty}, {Fender}, {Fridriksson}, {Remillard}, \& {Schulz}}]{Homan_2016}
{Homan}, J., {Neilsen}, J., {Allen}, J.~L., {et~al.} 2016, \apjl, 830, L5

\bibitem[{{Homan} {et~al.}(2019){Homan}, {Neilsen}, {Steiner}, {Remillard}, {Altamirano}, {Gendreau}, \& {Arzoumanian}}]{Homan_2019_Atel12742}
{Homan}, J., {Neilsen}, J., {Steiner}, J., {et~al.} 2019, The Astronomer's Telegram, 12742, 1

\bibitem[{{Homan} {et~al.}(2001){Homan}, {Wijnands}, {van der Klis}, {Belloni}, {van Paradijs}, {Klein-Wolt}, {Fender}, \& {M{\'e}ndez}}]{Homan_2001}
{Homan}, J., {Wijnands}, R., {van der Klis}, M., {et~al.} 2001, \apjs, 132, 377

\bibitem[{{Houck} \& {Denicola}(2000)}]{Houck_2000}
{Houck}, J.~C. \& {Denicola}, L.~A. 2000, in Astronomical Society of the Pacific Conference Series, Vol. 216, Astronomical Data Analysis Software and Systems IX, ed. N.~{Manset}, C.~{Veillet}, \& D.~{Crabtree}, 591

\bibitem[{{House}(1969)}]{House_1969}
{House}, L.~L. 1969, \apjs, 18, 21

\bibitem[{{Huppenkothen} {et~al.}(2019{\natexlab{a}}){Huppenkothen}, {Bachetti}, {Stevens}, {Migliari}, {Balm}, {Hammad}, {Khan}, {Mishra}, {Rashid}, {Sharma}, {Ribeiro}, \& {Blanco}}]{Huppenkothen_2019_stingrayb}
{Huppenkothen}, D., {Bachetti}, M., {Stevens}, A., {et~al.} 2019{\natexlab{a}}, The Journal of Open Source Software, 4, 1393

\bibitem[{{Huppenkothen} {et~al.}(2019{\natexlab{b}}){Huppenkothen}, {Bachetti}, {Stevens}, {Migliari}, {Balm}, {Hammad}, {Khan}, {Mishra}, {Rashid}, {Sharma}, {Martinez Ribeiro}, \& {Valles Blanco}}]{Huppenkothen_2019_stingraya}
{Huppenkothen}, D., {Bachetti}, M., {Stevens}, A.~L., {et~al.} 2019{\natexlab{b}}, \apj, 881, 39

\bibitem[{{Ingram} \& {Done}(2011)}]{Ingram_2011}
{Ingram}, A. \& {Done}, C. 2011, \mnras, 415, 2323

\bibitem[{{Ingram} {et~al.}(2009){Ingram}, {Done}, \& {Fragile}}]{Ingram_2009}
{Ingram}, A., {Done}, C., \& {Fragile}, P.~C. 2009, \mnras, 397, L101

\bibitem[{{Kaastra} \& {Bleeker}(2016)}]{Kaastra_2016}
{Kaastra}, J.~S. \& {Bleeker}, J.~A.~M. 2016, \aap, 587, A151

\bibitem[{{Kalemci} {et~al.}(2016){Kalemci}, {Begelman}, {Maccarone}, {Din{\c{c}}er}, {Russell}, {Bailyn}, \& {Tomsick}}]{Kalemci_2016}
{Kalemci}, E., {Begelman}, M.~C., {Maccarone}, T.~J., {et~al.} 2016, \mnras, 463, 615

\bibitem[{{Kallman} {et~al.}(2004){Kallman}, {Palmeri}, {Bautista}, {Mendoza}, \& {Krolik}}]{Kallman_2004}
{Kallman}, T.~R., {Palmeri}, P., {Bautista}, M.~A., {Mendoza}, C., \& {Krolik}, J.~H. 2004, \apjs, 155, 675

\bibitem[{{King} {et~al.}(2015){King}, {Miller}, {Raymond}, {Reynolds}, \& {Morningstar}}]{King_2015}
{King}, A.~L., {Miller}, J.~M., {Raymond}, J., {Reynolds}, M.~T., \& {Morningstar}, W. 2015, \apjl, 813, L37

\bibitem[{{Klein-Wolt} {et~al.}(2002){Klein-Wolt}, {Fender}, {Pooley}, {Belloni}, {Migliari}, {Morgan}, \& {van der Klis}}]{Klein-wolt_2002}
{Klein-Wolt}, M., {Fender}, R.~P., {Pooley}, G.~G., {et~al.} 2002, \mnras, 331, 745

\bibitem[{{Koljonen} {et~al.}(2019){Koljonen}, {Vera}, {Lahteenmaki}, \& {Tornikoski}}]{Koljonen_2019_Atel12839}
{Koljonen}, K., {Vera}, R., {Lahteenmaki}, A., \& {Tornikoski}, M. 2019, The Astronomer's Telegram, 12839, 1

\bibitem[{{Koljonen} \& {Hovatta}(2021)}]{Koljonen_2021}
{Koljonen}, K.~I.~I. \& {Hovatta}, T. 2021, \aap, 647, A173

\bibitem[{{Koljonen} \& {Tomsick}(2020)}]{Koljonen_2020}
{Koljonen}, K.~I.~I. \& {Tomsick}, J.~A. 2020, \aap, 639, A13

\bibitem[{{Kong} {et~al.}(2024){Kong}, {Ji}, {Santangelo}, {Zhou}, {Shui}, \& {Zhang}}]{Kong_2024}
{Kong}, L.-D., {Ji}, L., {Santangelo}, A., {et~al.} 2024, \aap, 686, A211

\bibitem[{{Kong} {et~al.}(2021){Kong}, {Zhang}, {Chen}, {Zhang}, {Ji}, {Wang}, {Tao}, {Ge}, {Liu}, {Song}, {Lu}, {Qu}, {Li}, {Xu}, {Cao}, {Chen}, {Bu}, {Cai}, {Chang}, {Chen}, {Chen}, {Chen}, {Cui}, {Du}, {Gao}, {Gao}, {Gao}, {Gu}, {Guan}, {Guo}, {Han}, {Huang}, {Huo}, {Jia}, {Jiang}, {Jin}, {Li}, {Li}, {Li}, {Li}, {Li}, {Li}, {Li}, {Li}, {Liang}, {Liao}, {Liu}, {Liu}, {Liu}, {Liu}, {Lu}, {Luo}, {Luo}, {Ma}, {Ma}, {Meng}, {Nang}, {Nie}, {Ou}, {Ren}, {Sai}, {Song}, {Sun}, {Tan}, {Tuo}, {Wang}, {Wang}, {Wang}, {Wang}, {Wen}, {Wu}, {Wu}, {Wu}, {Xiao}, {Xiao}, {Xiong}, {Yang}, {Yang}, {Yang}, {Yang}, {Yi}, {Yin}, {You}, {Zhang}, {Zhang}, {Zhang}, {Zhang}, {Zhang}, {Zhang}, {Zhang}, {Zhang}, {Zhao}, {Zhao}, {Zheng}, {Zheng}, \& {Zhou}}]{Kong_2021}
{Kong}, L.~D., {Zhang}, S., {Chen}, Y.~P., {et~al.} 2021, \apjl, 906, L2

\bibitem[{{Krawczynski} {et~al.}(2022){Krawczynski}, {Muleri}, {Dov{\v{c}}iak}, {Veledina}, {Rodriguez Cavero}, {Svoboda}, {Ingram}, {Matt}, {Garcia}, {Loktev}, {Negro}, {Poutanen}, {Kitaguchi}, {Podgorn{\'y}}, {Rankin}, {Zhang}, {Berdyugin}, {Berdyugina}, {Bianchi}, {Blinov}, {Capitanio}, {Di Lalla}, {Draghis}, {Fabiani}, {Kagitani}, {Kravtsov}, {Kiehlmann}, {Latronico}, {Lutovinov}, {Mandarakas}, {Marin}, {Marinucci}, {Miller}, {Mizuno}, {Molkov}, {Omodei}, {Petrucci}, {Ratheesh}, {Sakanoi}, {Semena}, {Skalidis}, {Soffitta}, {Tennant}, {Thalhammer}, {Tombesi}, {Weisskopf}, {Wilms}, {Zhang}, {Agudo}, {Antonelli}, {Bachetti}, {Baldini}, {Baumgartner}, {Bellazzini}, {Bongiorno}, {Bonino}, {Brez}, {Bucciantini}, {Castellano}, {Cavazzuti}, {Ciprini}, {Costa}, {De Rosa}, {Del Monte}, {Di Gesu}, {Di Marco}, {Donnarumma}, {Doroshenko}, {Ehlert}, {Enoto}, {Evangelista}, {Ferrazzoli}, {Gunji}, {Hayashida}, {Heyl}, {Iwakiri}, {Jorstad}, {Karas}, {Kolodziejczak}, {La Monaca}, {Liodakis}, {Maldera}, {Manfreda},
  {Marscher}, {Marshall}, {Mitsuishi}, {Ng}, {O{\textquoteright}Dell}, {Oppedisano}, {Papitto}, {Pavlov}, {Peirson}, {Perri}, {Pesce-Rollins}, {Pilia}, {Possenti}, {Puccetti}, {Ramsey}, {Romani}, {Sgr{\`o}}, {Slane}, {Spandre}, {Tamagawa}, {Tavecchio}, {Taverna}, {Tawara}, {Thomas}, {Trois}, {Tsygankov}, {Turolla}, {Vink}, {Wu}, {Xie}, \& {Zane}}]{Krawczynski_2022}
{Krawczynski}, H., {Muleri}, F., {Dov{\v{c}}iak}, M., {et~al.} 2022, Science, 378, 650

\bibitem[{{Kubota} {et~al.}(2007){Kubota}, {Dotani}, {Cottam}, {Kotani}, {Done}, {Ueda}, {Fabian}, {Yasuda}, {Takahashi}, {Fukazawa}, {Yamaoka}, {Makishima}, {Yamada}, {Kohmura}, \& {Angelini}}]{Kubota_2007}
{Kubota}, A., {Dotani}, T., {Cottam}, J., {et~al.} 2007, \pasj, 59, 185

\bibitem[{{Lee} {et~al.}(2002){Lee}, {Reynolds}, {Remillard}, {Schulz}, {Blackman}, \& {Fabian}}]{Lee_2002}
{Lee}, J.~C., {Reynolds}, C.~S., {Remillard}, R., {et~al.} 2002, \apj, 567, 1102

\bibitem[{{Li} {et~al.}(2020){Li}, {Li}, {Tan}, {Yang}, {Ge}, {Zhang}, {Tuo}, {Wu}, {Liao}, {Zhang}, {Song}, {Zhang}, {Qu}, {Zhang}, {Lu}, {Xu}, {Liu}, {Cao}, {Chen}, {Nie}, {Zhao}, \& {Li}}]{Li_2020_flight}
{Li}, X., {Li}, X., {Tan}, Y., {et~al.} 2020, Journal of High Energy Astrophysics, 27, 64

\bibitem[{{Liao} {et~al.}(2020{\natexlab{a}}){Liao}, {Zhang}, {Chen}, {Zhang}, {Jin}, {Chang}, {Chen}, {Ge}, {Guo}, {Li}, {Li}, {Lu}, {Lu}, {Nie}, {Song}, {Yang}, {You}, {Zhao}, \& {Zhang}}]{Liao_2020_lebackground}
{Liao}, J.-Y., {Zhang}, S., {Chen}, Y., {et~al.} 2020{\natexlab{a}}, Journal of High Energy Astrophysics, 27, 24

\bibitem[{{Liao} {et~al.}(2020{\natexlab{b}}){Liao}, {Zhang}, {Lu}, {Zhang}, {Li}, {Chang}, {Chen}, {Ge}, {Guo}, {Huang}, {Jin}, {Li}, {Li}, {Li}, {Liu}, {Lu}, {Nie}, {Song}, {Wang}, {You}, {Zhang}, {Zhao}, \& {Zhang}}]{Liao_2020_hebackground}
{Liao}, J.-Y., {Zhang}, S., {Lu}, X.-F., {et~al.} 2020{\natexlab{b}}, Journal of High Energy Astrophysics, 27, 14

\bibitem[{{Liu} {et~al.}(2020){Liu}, {Zhang}, {Li}, {Lu}, {Chang}, {Li}, {Zhang}, {Jin}, {Yu}, {Zhang}, {Fu}, {Chen}, {Ji}, {Xu}, {Deng}, {Shang}, {Liu}, {Lu}, {Zhang}, {Dong}, {Li}, {Wu}, {Li}, {Wang}, {Wu}, {Zhang}, {Zhang}, {Xiong}, {Liu}, {Zhang}, {Liu}, {Yang}, \& {Zhang}}]{Liu_2020_he}
{Liu}, C., {Zhang}, Y., {Li}, X., {et~al.} 2020, Science China Physics, Mechanics, and Astronomy, 63, 249503

\bibitem[{{Makishima} {et~al.}(1986){Makishima}, {Maejima}, {Mitsuda}, {Bradt}, {Remillard}, {Tuohy}, {Hoshi}, \& {Nakagawa}}]{Makishima_1986}
{Makishima}, K., {Maejima}, Y., {Mitsuda}, K., {et~al.} 1986, \apj, 308, 635

\bibitem[{{Markoff} {et~al.}(2005){Markoff}, {Nowak}, \& {Wilms}}]{Markoff_2005}
{Markoff}, S., {Nowak}, M.~A., \& {Wilms}, J. 2005, \apj, 635, 1203

\bibitem[{{Martocchia} {et~al.}(2006){Martocchia}, {Matt}, {Belloni}, {Feroci}, {Karas}, \& {Ponti}}]{Martocchia_2006}
{Martocchia}, A., {Matt}, G., {Belloni}, T., {et~al.} 2006, \aap, 448, 677

\bibitem[{{Matsuoka} {et~al.}(2009){Matsuoka}, {Kawasaki}, {Ueno}, {Tomida}, {Kohama}, {Suzuki}, {Adachi}, {Ishikawa}, {Mihara}, {Sugizaki}, {Isobe}, {Nakagawa}, {Tsunemi}, {Miyata}, {Kawai}, {Kataoka}, {Morii}, {Yoshida}, {Negoro}, {Nakajima}, {Ueda}, {Chujo}, {Yamaoka}, {Yamazaki}, {Nakahira}, {You}, {Ishiwata}, {Miyoshi}, {Eguchi}, {Hiroi}, {Katayama}, \& {Ebisawa}}]{Matsuoka_2009}
{Matsuoka}, M., {Kawasaki}, K., {Ueno}, S., {et~al.} 2009, \pasj, 61, 999

\bibitem[{{Matt} {et~al.}(2003){Matt}, {Guainazzi}, \& {Maiolino}}]{Matt_2003}
{Matt}, G., {Guainazzi}, M., \& {Maiolino}, R. 2003, \mnras, 342, 422

\bibitem[{{Merloni} \& {Fabian}(2002)}]{Merloni_2002}
{Merloni}, A. \& {Fabian}, A.~C. 2002, \mnras, 332, 165

\bibitem[{{Miller} {et~al.}(2013){Miller}, {Parker}, {Fuerst}, {Bachetti}, {Harrison}, {Barret}, {Boggs}, {Chakrabarty}, {Christensen}, {Craig}, {Fabian}, {Grefenstette}, {Hailey}, {King}, {Stern}, {Tomsick}, {Walton}, \& {Zhang}}]{Miller_2013}
{Miller}, J.~M., {Parker}, M.~L., {Fuerst}, F., {et~al.} 2013, \apjl, 775, L45

\bibitem[{{Miller} {et~al.}(2006{\natexlab{a}}){Miller}, {Raymond}, {Fabian}, {Steeghs}, {Homan}, {Reynolds}, {van der Klis}, \& {Wijnands}}]{Miller_2006_Nature}
{Miller}, J.~M., {Raymond}, J., {Fabian}, A., {et~al.} 2006{\natexlab{a}}, \nat, 441, 953

\bibitem[{{Miller} {et~al.}(2016){Miller}, {Raymond}, {Fabian}, {Gallo}, {Kaastra}, {Kallman}, {King}, {Proga}, {Reynolds}, \& {Zoghbi}}]{Miller_2016}
{Miller}, J.~M., {Raymond}, J., {Fabian}, A.~C., {et~al.} 2016, \apjl, 821, L9

\bibitem[{{Miller} {et~al.}(2006{\natexlab{b}}){Miller}, {Raymond}, {Homan}, {Fabian}, {Steeghs}, {Wijnands}, {Rupen}, {Charles}, {van der Klis}, \& {Lewin}}]{Miller_2006_ApJ}
{Miller}, J.~M., {Raymond}, J., {Homan}, J., {et~al.} 2006{\natexlab{b}}, \apj, 646, 394

\bibitem[{{Miller} {et~al.}(2020){Miller}, {Zoghbi}, {Raymond}, {Balakrishnan}, {Brenneman}, {Cackett}, {Draghis}, {Fabian}, {Gallo}, {Kaastra}, {Kallman}, {Kammoun}, {Motta}, {Proga}, {Reynolds}, \& {Trueba}}]{Miller_2020}
{Miller}, J.~M., {Zoghbi}, A., {Raymond}, J., {et~al.} 2020, \apj, 904, 30

\bibitem[{{Mirabel} \& {Rodr{\'\i}guez}(1994)}]{Mirabel_1994}
{Mirabel}, I.~F. \& {Rodr{\'\i}guez}, L.~F. 1994, \nat, 371, 46

\bibitem[{{Mirzaev} {et~al.}(2024){Mirzaev}, {Riaz}, {Abdikamalov}, {Bambi}, {Dauser}, {Garcia}, {Jiang}, {Liu}, \& {Shashank}}]{Mirzaev_2024}
{Mirzaev}, T., {Riaz}, S., {Abdikamalov}, A.~B., {et~al.} 2024, \apj, 965, 66

\bibitem[{{Mitsuda} {et~al.}(1984){Mitsuda}, {Inoue}, {Koyama}, {Makishima}, {Matsuoka}, {Ogawara}, {Shibazaki}, {Suzuki}, {Tanaka}, \& {Hirano}}]{Mitsuda_1984}
{Mitsuda}, K., {Inoue}, H., {Koyama}, K., {et~al.} 1984, \pasj, 36, 741

\bibitem[{{Morgan} {et~al.}(1997){Morgan}, {Remillard}, \& {Greiner}}]{Morgan_1997}
{Morgan}, E.~H., {Remillard}, R.~A., \& {Greiner}, J. 1997, \apj, 482, 993

\bibitem[{{Motta} {et~al.}(2011){Motta}, {Mu{\~n}oz-Darias}, {Casella}, {Belloni}, \& {Homan}}]{Motta_2011}
{Motta}, S., {Mu{\~n}oz-Darias}, T., {Casella}, P., {Belloni}, T., \& {Homan}, J. 2011, \mnras, 418, 2292

\bibitem[{{Motta} {et~al.}(2019){Motta}, {Williams}, {Fender}, {Titterington}, {Green}, \& {Perrott}}]{Motta_2019_Atel12773}
{Motta}, S., {Williams}, D., {Fender}, R., {et~al.} 2019, The Astronomer's Telegram, 12773, 1

\bibitem[{{Motta} {et~al.}(2021){Motta}, {Kajava}, {Giustini}, {Williams}, {Del Santo}, {Fender}, {Green}, {Heywood}, {Rhodes}, {Segreto}, {Sivakoff}, \& {Woudt}}]{Motta_2021}
{Motta}, S.~E., {Kajava}, J.~J.~E., {Giustini}, M., {et~al.} 2021, \mnras, 503, 152

\bibitem[{{Motta} {et~al.}(2017{\natexlab{a}}){Motta}, {Kajava}, {S{\'a}nchez-Fern{\'a}ndez}, {Beardmore}, {Sanna}, {Page}, {Fender}, {Altamirano}, {Charles}, {Giustini}, {Knigge}, {Kuulkers}, {Oates}, \& {Osborne}}]{Motta_2017_Oct}
{Motta}, S.~E., {Kajava}, J.~J.~E., {S{\'a}nchez-Fern{\'a}ndez}, C., {et~al.} 2017{\natexlab{a}}, \mnras, 471, 1797

\bibitem[{{Motta} {et~al.}(2017{\natexlab{b}}){Motta}, {Kajava}, {S{\'a}nchez-Fern{\'a}ndez}, {Giustini}, \& {Kuulkers}}]{Motta_2017_Jun}
{Motta}, S.~E., {Kajava}, J.~J.~E., {S{\'a}nchez-Fern{\'a}ndez}, C., {Giustini}, M., \& {Kuulkers}, E. 2017{\natexlab{b}}, \mnras, 468, 981

\bibitem[{{Narayan} \& {Yi}(1995)}]{Narayan_1995}
{Narayan}, R. \& {Yi}, I. 1995, \apj, 444, 231

\bibitem[{{Negoro} {et~al.}(2018){Negoro}, {Tachibana}, {Kawai}, {Yamaoka}, {Ueda}, {Nakajima}, {Sakamaki}, {Maruyama}, {Mihara}, {Nakahira}, {Yatabe}, {Takao}, {Matsuoka}, {Sakamoto}, {Serino}, {Sugita}, {Kawakubo}, {Hashimoto}, {Yoshida}, {Sugizaki}, {Morita}, {Ueno}, {Tomida}, {Ishikawa}, {Sugawara}, {Isobe}, {Shimomukai}, {Tanimoto}, {Morita}, {Yamada}, {Tsuboi}, {Iwakiri}, {Sasaki}, {Kawai}, {Sato}, {Tsunemi}, {Yoneyama}, {Yamauchi}, {Hidaka}, {Iwahori}, {Kawamuro}, \& {Shidatsu}}]{Negoro_2018_Atel11828}
{Negoro}, H., {Tachibana}, Y., {Kawai}, N., {et~al.} 2018, The Astronomer's Telegram, 11828, 1

\bibitem[{{Neilsen} {et~al.}(2018){Neilsen}, {Cackett}, {Remillard}, {Homan}, {Steiner}, {Gendreau}, {Arzoumanian}, {Prigozhin}, {LaMarr}, {Doty}, {Eikenberry}, {Tombesi}, {Ludlam}, {Kara}, {Altamirano}, \& {Fabian}}]{Neilsen_2018}
{Neilsen}, J., {Cackett}, E., {Remillard}, R.~A., {et~al.} 2018, \apjl, 860, L19

\bibitem[{{Neilsen} {et~al.}(2020){Neilsen}, {Homan}, {Steiner}, {Marcel}, {Cackett}, {Remillard}, \& {Gendreau}}]{Neilsen_2020}
{Neilsen}, J., {Homan}, J., {Steiner}, J.~F., {et~al.} 2020, \apj, 902, 152

\bibitem[{{Neilsen} \& {Lee}(2009)}]{Neilsen_2009}
{Neilsen}, J. \& {Lee}, J.~C. 2009, \nat, 458, 481

\bibitem[{{Nowak}(2000)}]{Nowak_2000}
{Nowak}, M.~A. 2000, \mnras, 318, 361

\bibitem[{{Petrucci} {et~al.}(2001){Petrucci}, {Merloni}, {Fabian}, {Haardt}, \& {Gallo}}]{Petrucci_2001}
{Petrucci}, P.~O., {Merloni}, A., {Fabian}, A., {Haardt}, F., \& {Gallo}, E. 2001, \mnras, 328, 501

\bibitem[{{Ponti} {et~al.}(2012){Ponti}, {Fender}, {Begelman}, {Dunn}, {Neilsen}, \& {Coriat}}]{Ponti_2012}
{Ponti}, G., {Fender}, R.~P., {Begelman}, M.~C., {et~al.} 2012, \mnras, 422, L11

\bibitem[{{Proga} \& {Kallman}(2002)}]{Proga_2002}
{Proga}, D. \& {Kallman}, T.~R. 2002, \apj, 565, 455

\bibitem[{{Reeves} {et~al.}(2008){Reeves}, {Done}, {Pounds}, {Terashima}, {Hayashida}, {Anabuki}, {Uchino}, \& {Turner}}]{Reeves_2008}
{Reeves}, J., {Done}, C., {Pounds}, K., {et~al.} 2008, \mnras, 385, L108

\bibitem[{{Reid} {et~al.}(2014){Reid}, {McClintock}, {Steiner}, {Steeghs}, {Remillard}, {Dhawan}, \& {Narayan}}]{Reid_2014}
{Reid}, M.~J., {McClintock}, J.~E., {Steiner}, J.~F., {et~al.} 2014, \apj, 796, 2

\bibitem[{{Remillard} {et~al.}(2022){Remillard}, {Loewenstein}, {Steiner}, {Prigozhin}, {LaMarr}, {Enoto}, {Gendreau}, {Arzoumanian}, {Markwardt}, {Basak}, {Stevens}, {Ray}, {Altamirano}, \& {Buisson}}]{Remillard_2022}
{Remillard}, R.~A., {Loewenstein}, M., {Steiner}, J.~F., {et~al.} 2022, \aj, 163, 130

\bibitem[{{Remillard} \& {McClintock}(2006)}]{Remillard_2006}
{Remillard}, R.~A. \& {McClintock}, J.~E. 2006, \araa, 44, 49

\bibitem[{{Remillard} {et~al.}(1999){Remillard}, {Morgan}, {McClintock}, {Bailyn}, \& {Orosz}}]{Remillard_1999}
{Remillard}, R.~A., {Morgan}, E.~H., {McClintock}, J.~E., {Bailyn}, C.~D., \& {Orosz}, J.~A. 1999, \apj, 522, 397

\bibitem[{{Shaw} {et~al.}(2022){Shaw}, {Miller}, {Grinberg}, {Buisson}, {Heinke}, {Plotkin}, {Tomsick}, {Bahramian}, {Gandhi}, \& {Sivakoff}}]{Shaw_2022}
{Shaw}, A.~W., {Miller}, J.~M., {Grinberg}, V., {et~al.} 2022, \mnras, 516, 124

\bibitem[{{Shi} {et~al.}(2023){Shi}, {Wu}, {Yan}, {Lyu}, \& {Liu}}]{Shi_2023}
{Shi}, Z., {Wu}, Q., {Yan}, Z., {Lyu}, B., \& {Liu}, H. 2023, \mnras, 525, 1431

\bibitem[{{Shidatsu} {et~al.}(2013){Shidatsu}, {Ueda}, {Nakahira}, {Done}, {Morihana}, {Sugizaki}, {Mihara}, {Hori}, {Negoro}, {Kawai}, {Yamaoka}, {Ebisawa}, {Matsuoka}, {Serino}, {Yoshikawa}, {Nagayama}, \& {Matsunaga}}]{Shidatsu_2013}
{Shidatsu}, M., {Ueda}, Y., {Nakahira}, S., {et~al.} 2013, \apj, 779, 26

\bibitem[{{Shreeram} \& {Ingram}(2020)}]{Shreeram_2020}
{Shreeram}, S. \& {Ingram}, A. 2020, \mnras, 492, 405

\bibitem[{{Soleri} {et~al.}(2008){Soleri}, {Belloni}, \& {Casella}}]{Soleri_2008}
{Soleri}, P., {Belloni}, T., \& {Casella}, P. 2008, \mnras, 383, 1089

\bibitem[{{Steiner} {et~al.}(2016){Steiner}, {Remillard}, {Garc{\'\i}a}, \& {McClintock}}]{Steiner_2016}
{Steiner}, J.~F., {Remillard}, R.~A., {Garc{\'\i}a}, J.~A., \& {McClintock}, J.~E. 2016, \apjl, 829, L22

\bibitem[{{Tagger} \& {Pellat}(1999)}]{Tagger_1999}
{Tagger}, M. \& {Pellat}, R. 1999, \aap, 349, 1003

\bibitem[{{Tarter} {et~al.}(1969){Tarter}, {Tucker}, \& {Salpeter}}]{Tarter_1969}
{Tarter}, C.~B., {Tucker}, W.~H., \& {Salpeter}, E.~E. 1969, \apj, 156, 943

\bibitem[{{Tomsick} \& {Kaaret}(2000)}]{Tomsick_2000}
{Tomsick}, J.~A. \& {Kaaret}, P. 2000, \apj, 537, 448

\bibitem[{{Ueda} {et~al.}(2009){Ueda}, {Yamaoka}, \& {Remillard}}]{Ueda_2009}
{Ueda}, Y., {Yamaoka}, K., \& {Remillard}, R. 2009, \apj, 695, 888

\bibitem[{{Veledina} {et~al.}(2023){Veledina}, {Muleri}, {Dov{\v{c}}iak}, {Poutanen}, {Ratheesh}, {Capitanio}, {Matt}, {Soffitta}, {Tennant}, {Negro}, {Kaaret}, {Costa}, {Ingram}, {Svoboda}, {Krawczynski}, {Bianchi}, {Steiner}, {Garc{\'\i}a}, {Kravtsov}, {Nitindala}, {Ewing}, {Mastroserio}, {Marinucci}, {Ursini}, {Tombesi}, {Tsygankov}, {Yang}, {Weisskopf}, {Trushkin}, {Egron}, {Iacolina}, {Pilia}, {Marra}, {Miku{\v{s}}incov{\'a}}, {Nathan}, {Parra}, {Petrucci}, {Podgorn{\'y}}, {Tugliani}, {Zane}, {Zhang}, {Agudo}, {Antonelli}, {Bachetti}, {Baldini}, {Baumgartner}, {Bellazzini}, {Bongiorno}, {Bonino}, {Brez}, {Bucciantini}, {Castellano}, {Cavazzuti}, {Chen}, {Ciprini}, {De Rosa}, {Del Monte}, {Di Gesu}, {Di Lalla}, {Di Marco}, {Donnarumma}, {Doroshenko}, {Ehlert}, {Enoto}, {Evangelista}, {Fabiani}, {Ferrazzoli}, {Gunji}, {Hayashida}, {Heyl}, {Iwakiri}, {Jorstad}, {Karas}, {Kislat}, {Kitaguchi}, {Kolodziejczak}, {La Monaca}, {Latronico}, {Liodakis}, {Maldera}, {Manfreda}, {Marin}, {Marscher}, {Marshall},
  {Massaro}, {Mitsuishi}, {Mizuno}, {Ng}, {O'Dell}, {Omodei}, {Oppedisano}, {Papitto}, {Pavlov}, {Peirson}, {Perri}, {Pesce-Rollins}, {Possenti}, {Puccetti}, {Ramsey}, {Rankin}, {Roberts}, {Romani}, {Sgr{\`o}}, {Slane}, {Spandre}, {Swartz}, {Tamagawa}, {Tavecchio}, {Taverna}, {Tawara}, {Thomas}, {Trois}, {Turolla}, {Vink}, {Wu}, \& {Xie}}]{Veledina_2023}
{Veledina}, A., {Muleri}, F., {Dov{\v{c}}iak}, M., {et~al.} 2023, \apjl, 958, L16

\bibitem[{{Verner} {et~al.}(1996){Verner}, {Ferland}, {Korista}, \& {Yakovlev}}]{Verner_1996}
{Verner}, D.~A., {Ferland}, G.~J., {Korista}, K.~T., \& {Yakovlev}, D.~G. 1996, \apj, 465, 487

\bibitem[{{Wang} {et~al.}(2021){Wang}, {Ji}, {Garc{\'\i}a}, {Dauser}, {M{\'e}ndez}, {Mao}, {Tao}, {Altamirano}, {Maggi}, {Zhang}, {Ge}, {Zhang}, {Qu}, {Zhang}, {Ma}, {Lu}, {Li}, {Huang}, {Zheng}, {Chang}, {Tuo}, {Song}, {Xu}, {Chen}, {Liu}, {Bu}, {Cai}, {Cao}, {Chen}, {Chen}, {Chen}, {Cui}, {Du}, {Gao}, {Gu}, {Guan}, {Guo}, {Han}, {Huo}, {Jia}, {Jiang}, {Jin}, {Kong}, {Li}, {Li}, {Li}, {Li}, {Li}, {Li}, {Li}, {Li}, {Liang}, {Liao}, {Liu}, {Liu}, {Lu}, {Luo}, {Luo}, {Meng}, {Nang}, {Nie}, {Ou}, {Sai}, {Shang}, {Song}, {Sun}, {Tan}, {Wang}, {Wang}, {Wang}, {Wen}, {Wu}, {Wu}, {Wu}, {Xiao}, {Xiao}, {Xiong}, {Yang}, {Yang}, {Yi}, {Yin}, {You}, {Zhang}, {Zhang}, {Zhang}, {Zhang}, {Zhang}, {Zhang}, {Zhao}, {Zhao}, \& {Zhou}}]{Wang_2021}
{Wang}, Y., {Ji}, L., {Garc{\'\i}a}, J.~A., {et~al.} 2021, \apj, 906, 11

\bibitem[{{Williams} {et~al.}(2020){Williams}, {Motta}, {Fender}, {Bright}, {Heywood}, {Tremou}, {Woudt}, {Buckley}, {Corbel}, {Coriat}, {Joseph}, {Rhodes}, {Sivakoff}, \& {van der Horst}}]{Williams_2020}
{Williams}, D.~R.~A., {Motta}, S.~E., {Fender}, R., {et~al.} 2020, \mnras, 491, L29

\bibitem[{{Wilms} {et~al.}(2000){Wilms}, {Allen}, \& {McCray}}]{Wilms_2000}
{Wilms}, J., {Allen}, A., \& {McCray}, R. 2000, \apj, 542, 914

\bibitem[{{Woods} {et~al.}(1996){Woods}, {Klein}, {Castor}, {McKee}, \& {Bell}}]{Woods_1996}
{Woods}, D.~T., {Klein}, R.~I., {Castor}, J.~I., {McKee}, C.~F., \& {Bell}, J.~B. 1996, \apj, 461, 767

\bibitem[{{Xu} {et~al.}(2018){Xu}, {Harrison}, {Kennea}, {Walton}, {Tomsick}, {Miller}, {Barret}, {Fabian}, {Forster}, {F{\"u}rst}, {Gandhi}, \& {Garc{\'\i}a}}]{Xu_2018}
{Xu}, Y., {Harrison}, F.~A., {Kennea}, J.~A., {et~al.} 2018, \apj, 865, 18

\bibitem[{{Zdziarski} {et~al.}(2004){Zdziarski}, {Gierli{\'n}ski}, {Miko{\l}ajewska}, {Wardzi{\'n}ski}, {Smith}, {Harmon}, \& {Kitamoto}}]{Zdziarski_2004}
{Zdziarski}, A.~A., {Gierli{\'n}ski}, M., {Miko{\l}ajewska}, J., {et~al.} 2004, \mnras, 351, 791

\bibitem[{{Zdziarski} {et~al.}(1996){Zdziarski}, {Johnson}, \& {Magdziarz}}]{Zdziarski_1996}
{Zdziarski}, A.~A., {Johnson}, W.~N., \& {Magdziarz}, P. 1996, \mnras, 283, 193

\bibitem[{{Zhang} {et~al.}(2014){Zhang}, {Lu}, {Zhang}, \& {Li}}]{Zhang_2014}
{Zhang}, S., {Lu}, F.~J., {Zhang}, S.~N., \& {Li}, T.~P. 2014, in Society of Photo-Optical Instrumentation Engineers (SPIE) Conference Series, Vol. 9144, Space Telescopes and Instrumentation 2014: Ultraviolet to Gamma Ray, ed. T.~{Takahashi}, J.-W.~A. {den Herder}, \& M.~{Bautz}, 914421

\bibitem[{{Zhang} {et~al.}(2020){Zhang}, {Li}, {Lu}, {Song}, {Xu}, {Liu}, {Chen}, {Cao}, {Bu}, {Chang}, {Chen}, {Chen}, {Chen}, {Chen}, {Chen}, {Cui}, {Cui}, {Deng}, {Dong}, {Du}, {Fu}, {Gao}, {Gao}, {Gao}, {Ge}, {Gu}, {Guan}, {Gungor}, {Guo}, {Han}, {Hu}, {Huang}, {Huo}, {Jia}, {Jiang}, {Jiang}, {Jin}, {Jin}, {Li}, {Li}, {Li}, {Li}, {Li}, {Li}, {Li}, {Li}, {Li}, {Li}, {Li}, {Liang}, {Liao}, {Liu}, {Liu}, {Liu}, {Liu}, {Liu}, {Liu}, {Lu}, {Lu}, {Luo}, {Ma}, {Meng}, {Nang}, {Nie}, {Ou}, {Qu}, {Sai}, {Shang}, {Shen}, {Sun}, {Tan}, {Tao}, {Tuo}, {Wang}, {Wang}, {Wang}, {Wang}, {Wang}, {Wang}, {Wang}, {Wen}, {Wu}, {Wu}, {Wu}, {Xiao}, {Xiong}, {Yan}, {Yang}, {Yang}, {Yang}, {Yi}, {Yuan}, {Zhang}, {Zhang}, {Zhang}, {Zhang}, {Zhang}, {Zhang}, {Zhang}, {Zhang}, {Zhang}, {Zhang}, {Zhang}, {Zhang}, {Zhang}, {Zhang}, {Zhang}, {Zhang}, {Zhang}, {Zhang}, {Zhang}, {Zhang}, {Zhao}, {Zhao}, {Zheng}, {Zhou}, {Zhu}, {Zhu}, {Zhuang}, \& {Insight-HXMT Team}}]{Zhang_2020}
{Zhang}, S.-N., {Li}, T., {Lu}, F., {et~al.} 2020, Science China Physics, Mechanics, and Astronomy, 63, 249502

\bibitem[{{Zhang} {et~al.}(1995){Zhang}, {Jahoda}, {Swank}, {Morgan}, \& {Giles}}]{Zhang_1995}
{Zhang}, W., {Jahoda}, K., {Swank}, J.~H., {Morgan}, E.~H., \& {Giles}, A.~B. 1995, \apj, 449, 930

\bibitem[{{Zhou} {et~al.}(2022){Zhou}, {Grinberg}, {Bu}, {Santangelo}, {Cangemi}, {Diez}, {K{\"o}nig}, {Ji}, {Nowak}, {Pottschmidt}, {Rodriguez}, {Wilms}, {Zhang}, {Qu}, \& {Zhang}}]{Zhou_2022}
{Zhou}, M., {Grinberg}, V., {Bu}, Q.~C., {et~al.} 2022, \aap, 666, A172

\bibitem[{{Zoghbi} {et~al.}(2016){Zoghbi}, {Miller}, {King}, {Miller}, {Proga}, {Kallman}, {Fabian}, {Harrison}, {Kaastra}, {Raymond}, {Reynolds}, {Boggs}, {Christensen}, {Craig}, {Hailey}, {Stern}, \& {Zhang}}]{Zoghbi_2016}
{Zoghbi}, A., {Miller}, J.~M., {King}, A.~L., {et~al.} 2016, \apj, 833, 165

\bibitem[{{{\.Z}ycki} {et~al.}(1999){{\.Z}ycki}, {Done}, \& {Smith}}]{Zycki_1999}
{{\.Z}ycki}, P.~T., {Done}, C., \& {Smith}, D.~A. 1999, \mnras, 309, 561

\end{thebibliography}

\end{document}